\documentclass[aps,prfluids,showpacs,notitlepage,longbibliography]{revtex4-1}
\usepackage{epsfig,graphics,amssymb,amsmath,subeqnarray,color,bm,bbm}
\usepackage[toc,page]{appendix}
\usepackage{mathtools}
\usepackage{kpfonts}
\usepackage{xcolor}
\usepackage{float}
\usepackage{graphicx}
\usepackage[colorlinks=true,linkcolor=blue]{hyperref}

\begin{document}
\title{Higher-order force moments of active particles}
\author{Babak Nasouri}
\affiliation{Department of Mechanical Engineering, Institute of Applied Mathematics, University of British Columbia, Vancouver, British Columbia, Canada V6T 1Z4}
\author{Gwynn J. Elfring}
\email{gelfring@mech.ubc.ca}
\affiliation{Department of Mechanical Engineering, Institute of Applied Mathematics, University of British Columbia, Vancouver, British Columbia, Canada V6T 1Z4}
\date{\today}


\begin{abstract} 
Active particles moving through fluids generate disturbance flows due to their activity. For simplicity, the induced flow field is often modeled by the leading terms in a far-field approximation of the Stokes equations, whose coefficients are the force, torque and stresslet (zeroth and first-order force moments) of the active particle. This level of approximation is quite useful, but may also fail to predict more complex behaviors that are observed experimentally. In this study, to provide a better approximation, we evaluate the contribution of the second-order force moments to the flow field and, by reciprocal theorem, present explicit formulas for the stresslet dipole, rotlet dipole and potential dipole for an arbitrarily-shaped active particle. As examples of this method, we derive modified Fax\'en laws for active spherical particles and resolve higher-order moments for active rod-like particles.
\end{abstract}


\maketitle
\section{Introduction}
Self-propulsion is ubiquitous in nature. Be it at the macroscopic scale of flying birds or the microscopic scale of swimming bacteria, the motion of active matter results from converting internal or ambient energy into mechanical work without any external input \citep{ramaswamy2010}. At sufficiently small scales in viscous fluids, inertia is irrelevant and viscous dissipation dominates the motion of the fluid and active particles within them \citep{happel1981}. In the absence of inertia, `reciprocal' body distortions are ineffective as a propulsion mechanism, and so active particles must propel themselves by other means in this realm \citep{purcell1977}. There exist several techniques to achieve net locomotion in the low-Reynolds-number regime \citep{lauga2009a,lauga2011,nasouri2017}. For instance, microorganisms such as \textit{Paramecium} and \textit{Volvox} use small appendages called cilia to facilitate motion \citep{lodish2000}. Cilia generate thrust through a coordinated pattern of beating, which may arise from hydrodynamic \citep{niedermayer2008,brumley2012,nasouri2016} or basal \citep{quaranta2015,wan2016,klindt2017} interactions. Propulsion can also be achieved synthetically by chemically-active particles with asymmetric non-uniform surface properties \citep{anderson1989,golestanian2005,walther2013}. In both of these examples, the effect of surface activity is confined to a narrow region surrounding the particle and hence may be modelled using `apparent' slip velocities on the surface. This way, one can explicitly find the propulsion speed and thereby the disturbance flow field, in terms of prescribed (or measured) slip velocities \citep{elgeti2015}.

For the inertialess motion of sufficiently small particles in viscous fluids, the flow field is often approximated by far-field singularity solutions of the Stokes equations. To leading order, the flow field decays linearly by distance ($\sim1/|\mathbf{x}|$) and, at this level of approximation, the particle is replaced by a point force (i.e., zeroth-order force moment) that leads to flow \citep{kim1991}. The next-order correction to the flow field, which decays quadratically ($\sim1/|\mathbf{x}|^2$), can be expressed using a force-dipole (i.e., first-order force moment), which is decomposed into a torque (the antisymmetric part) and a stresslet (the symmetric part) \citep{batchelor1970}. In the absence of an external force, the over-damped motion of the particle has no net hydrodynamic force or torque and so the stresslet governs the leading-order flow field. The importance of the stresslet in characterizing the interactions of active particles \citep{guell1988,berke2008,lauga2016}, the rheology \citep{saintillan2009} and stability \citep{saintillan2013} of active suspensions, and the collective locomotion of bacteria \citep{dombrowski2004} is well documented. However, the stresslet term alone fails to explain behaviors such as the `dancing' of two \textit{Volvox} colonies when they are in proximity of one another \citep{drescher2009}, or the vortices induced due to the motion of \textit{C. reinhardtii} \citep{drescher2010,guasto2010}. An emerging picture is that modeling the motion using only terms up to the stresslet may limit understanding of how active particles interact with their environment and motivates investigation of higher-order force moments. In a recent study, \citet{ghose2014} showed that the swirling motion of an active spherical particle only appears in the flow field decaying as $\sim1/|\mathbf{x}|^3$ and $\sim1/|\mathbf{x}|^4$ and derived expressions for higher-order force moments of a sphere. In this study, we generalize their results by investigating the effects of higher-order force moments on an arbitrarily-shaped active particle, thereby extending recent general results for the stresslet term by \citet{lauga2016}. Using the boundary integral equations, we express the flow field around an active particle through a multipole expansion up to the contribution of the second-order force moments. We then provide explicit formulas for these force moments by exploiting the reciprocal theorem using a framework developed in \citep{elfring2017}.

The reciprocal theorem for low-Reynolds-number hydrodynamics has long been an avenue to simplify calculations in Stokes flow \citep{hinch1972,rallison1978,leal1979,happel1981}. Its application has ranged from the inertialess jet propulsion \citep{spagnolie2010}, boundary-driven channel flow \citep{michelin2015}, to Marangoni motion of a droplet covered with bulk-insoluble surfactants in a Poiseuille flow \citep{pak2014b}. In particular, \citet{stone1996} showed that the kinematics of an active particle can be determined explicitly, using the flow field induced by the rigid-body motion of a passive particle of the same instantaneous shape. Subsequently, the reciprocal theorem has been widely used to determine the kinematics of active particles both in Newtonian \citep{lauga2009a,elfring2015} and non-Newtonian fluids \citep{lauga09,pak12,lauga14,datt2015,datt2017}. This approach was recently extended to determine the stresslet of active particles \citep{lauga2016}. More recently, a general framework has been developed for finding the force moments (of any order) of an active particle in a Newtonian (or non-Newtonian) fluid \citep{elfring2017}. Following that approach, in this study, we provide formulas for calculating the force moments up to the second order, for any arbitrarily-shaped active particle.

The paper is organized as follows. In Sec.~\ref{multi}, we employ the boundary integral equation to describe the disturbance flow field caused by an active particle. Using an asymptotic expansion of the far-field flow, we show how the force moments contribute to the disturbance flow field. We then in Sec.~\ref{reciprocal}, use the reciprocal theorem to find general expressions for these force moments and, as examples, evaluate them explicitly for a spherical active particle, a generalized squirmer and an active slender rod.
\section{Multipole expansion} 
\label{multi}
\begin{figure}[H]
\begin{center}
\includegraphics[scale=0.55]{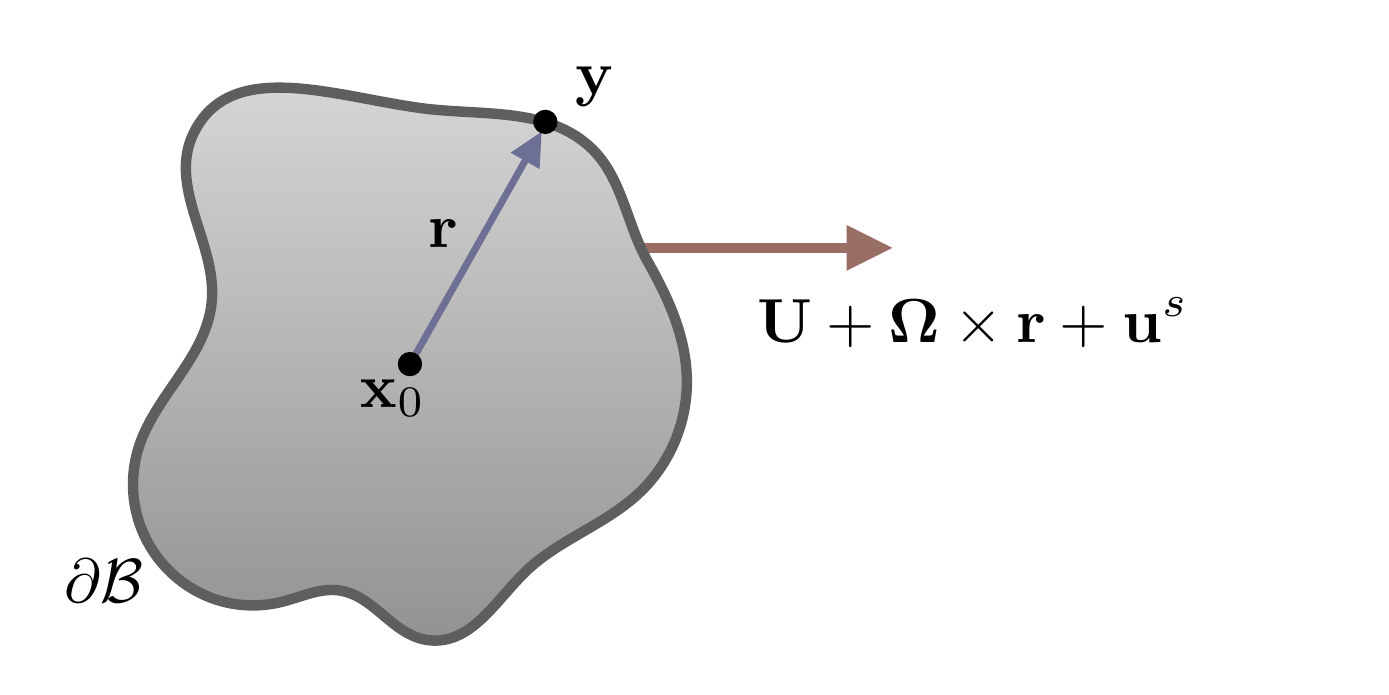}
\caption{Schematic representation of an active particle of arbitrary shape. A point on the particle surface, $\partial\mathcal{B}$, is denoted by $\mathbf{y}$ and $\mathbf{x}_0$ is a convenient reference point in the body. The instantaneous velocity of a point on $\partial\mathcal{B}$ is given by rigid-body translation $\mathbf{U}$, rigid-body rotation $\boldsymbol{\Omega}\times\mathbf{r}$ and surface slip velocity $\mathbf{u}^s$.}
\label{schematic}
\end{center}
\end{figure}

We consider a particle with boundary $\partial \mathcal{B}$ in an otherwise unbounded Newtonian fluid of viscosity $\mu$ and background flow field $\mathbf{u}^\infty$, as shown in Fig.~\ref{schematic}. Using the boundary integral equations, the disturbance flow field $\mathbf{u}^\prime=\mathbf{u}-\mathbf{u}^\infty$ can be expressed as a summation of \textit{single-layer} and \textit{double-layer} potentials \citep{pozrikidis1992,kim1991}
\begin{align}
\label{poz}
\mathbf{u}^\prime(\mathbf{x})=-\frac{1}{8\pi\mu}\int_{\partial \mathcal{B}} \mathbf{f}^{\prime}(\mathbf{y})\cdot\mathbf{J}(\mathbf{x}-\mathbf{y}) \text{d}S(\mathbf{y})-\frac{1}{8\pi}\int_{\partial \mathcal{B}}\mathbf{u}^\prime(\mathbf{y})\mathbf{n}(\mathbf{y}):\mathbf{K}(\mathbf{x}-\mathbf{y}) \text{d}S(\mathbf{y}),
\end{align}
where ${\mathbf{f}^{\prime}}=\mathbf{n}\cdot\boldsymbol{\sigma}^\prime$ is the traction of disturbance stress tensor $\boldsymbol{\sigma}^\prime$, $\mathbf{y}$ is the position that is integrated over the particle surface and $\mathbf{n}$ is the surface normal pointing into the fluid. Here $\mathbf{J({\mathbf{x}})}=\frac{\mathbf{I}}{|{\mathbf{x}}|}+\frac{{\mathbf{x}}{\mathbf{x}}}{|{\mathbf{x}}|^3}$ is the Green's function of Stokes equations (or the Oseen tensor) and $\mathbf{K}({\mathbf{x}})=\frac{-6{\mathbf{x}}{\mathbf{x}}{\mathbf{x}}}{|{\mathbf{x}}|^5}$ is its associated stress tensor. 
Expanding in $\mathbf{y}$ about a convenient point in the body $\mathbf{x}_0$ (for example the center of mass),
\begin{align}
\mathbf{J}(\mathbf{x}-\mathbf{y})&=\mathbf{J}(\mathbf{x}-\mathbf{x}_0) - \mathbf{r}\cdot \boldsymbol{\nabla}\mathbf{J}(\mathbf{x}-\mathbf{x}_0) +\frac{1}{2} \mathbf{r}\mathbf{r}: \boldsymbol{\nabla}\boldsymbol{\nabla}\mathbf{J}(\mathbf{x}-\mathbf{x}_0) +\cdots,\\
\mathbf{K}(\mathbf{x}-\mathbf{y})&=\mathbf{K}(\mathbf{x}-\mathbf{x}_0) - \mathbf{r}\cdot \boldsymbol{\nabla}\mathbf{K}(\mathbf{x}-\mathbf{x}_0)+\cdots,
\end{align}
where $\mathbf{r} = \mathbf{y}-\mathbf{x}_0$. Equation \eqref{poz} then takes the form
\begin{align}
\label{expansion}
\mathbf{u}^\prime(\mathbf{x})=&-\frac{1}{8\pi\mu}\left[ \left<\mathbf{f}^{\prime}\right>\cdot\mathbf{J}(\mathbf{x}-\mathbf{x}_0) - \left<\mathbf{f}^{\prime}\mathbf{r}\right>:\boldsymbol{\nabla}\mathbf{J}(\mathbf{x}-\mathbf{x}_0) +\frac{1}{2} \left<\mathbf{f}^{\prime}\mathbf{r}\mathbf{r}\right>\odot\boldsymbol{\nabla}\boldsymbol{\nabla}\mathbf{J}(\mathbf{x}-\mathbf{x}_0)+\cdots \right]\nonumber\\
&-\frac{1}{8\pi}\left[\left<\mathbf{u}^\prime\mathbf{n}\right>:\mathbf{K}(\mathbf{x}-\mathbf{x}_0) - \left<\mathbf{u}^\prime\mathbf{n}\mathbf{r}\right>\odot\boldsymbol{\nabla}\mathbf{K}(\mathbf{x}-\mathbf{x}_0)+\cdots \right].
\end{align}
For convenience, in this work, we denote the surface integral by $\int_{\partial\mathcal{B}}\cdots\text{d} S\equiv\left<\cdots\right>$.We also use $\odot$ to denote a $k-$fold contraction where $k=\text{min}\{a,b\}$ and $a$ and $b$ are the tensorial orders of the contracted tensors, e.g.,  $\left[ \left<\mathbf{u}^\prime\mathbf{n}\mathbf{r}\right>\odot\boldsymbol{\nabla}\mathbf{K}\right]_i=\left<{u}^\prime_j{n}_k\mathbf{r}_m\right>{\nabla}_m {K}_{kji}$. We define $\mathbf{u}^\prime(\mathbf{x})=\sum_{i=1} \mathbf{u}^{\prime\text{(i)}}(\mathbf{x})$, where $\mathbf{u}^{\prime\text{(i)}}(\mathbf{x})$ is the flow field that decays as $|\mathbf{x}|^{-i}$. At leading order, one can recognize the net hydrodynamic force $\mathbf{F}=\left<\mathbf{f}^{\prime}\right>$ as the zeroth-order force moment. The presence of particle at this order is represented by a point force of strength $-\mathbf{F}$ and the flow field is simply governed by a Stokeslet, 
\begin{align}
\mathbf{u}^{\prime\text{(1)}}(\mathbf{x})=-\frac{1}{8\pi\mu}\mathbf{F}\cdot\mathbf{J}(\mathbf{x}-\mathbf{x}_0).
\end{align}

To find the flow field decaying as $|\mathbf{x}|^{-2}$, it is useful to decompose $\mathbf{f}^\prime\mathbf{r}$ and $\mathbf{u}^\prime\mathbf{n}$, to their symmetric and antisymmetric parts as $\mathbf{f}^\prime \mathbf{r}=\frac{\mathbf{f}^\prime \mathbf{r}+\mathbf{r}\mathbf{f}^\prime}{2}+\frac{\mathbf{f}^\prime \mathbf{r}-\mathbf{r}\mathbf{f}^\prime}{2}$ and $\mathbf{u}^{\prime}\mathbf{n}=\frac{\mathbf{u}^{\prime}\mathbf{n}+\mathbf{n}\mathbf{u}^{\prime}}{2}+\frac{\mathbf{u}^\prime\mathbf{n}-\mathbf{n}\mathbf{u}^\prime}{2}$. Noting the symmetry of $\mathbf{K}(\mathbf{x})$ and also recalling
\begin{align}
{\nabla}_k J_{ij}=\frac{x_i \delta_{jk} + x_j \delta_{ik} - x_k \delta_{ij}}{(x_l x_l)^{\frac{3}{2}}}-\frac{3x_i x_j x_k}{(x_l x_l)^{\frac{5}{2}}},
\end{align}
we have
\begin{align}
\mathbf{u}^{\prime\text{(2)}}(\mathbf{x})&=-\frac{1}{8\pi\mu}\left[\left<\mathbf{r}\times\mathbf{f}^{\prime}\right>\cdot \mathbf{C}(\mathbf{x}-\mathbf{x}_0)-\frac{1}{2}\left<\frac{{\mathbf{f}^{\prime}\mathbf{r}}+{\mathbf{r}\mathbf{f}^{\prime}}}{2}-\frac{1}{3}(\mathbf{f}^{\prime}\cdot\mathbf{r})\mathbf{I}-\mu\left({\mathbf{u}^\prime\mathbf{n}}+{\mathbf{n} \mathbf{u}^\prime}\right)\right>:\mathbf{K}(\mathbf{x}-\mathbf{x}_0)\right],\nonumber\\
&=-\frac{1}{8\pi\mu}\left[\mathbf{L}\cdot\mathbf{C}(\mathbf{x}-\mathbf{x}_0) -\frac{1}{2}\mathbf{S}:\mathbf{K}(\mathbf{x}-\mathbf{x}_0)\right],
\end{align}
where $\mathbf{L}=\left<\mathbf{r}\times\mathbf{f}^{\prime}\right>$ is the antisymmetric first-order force moment, i.e. torque and $\mathbf{C}(\mathbf{x})=\frac{\mathbf{I}\times\mathbf{x}}{|\mathbf{x}|^3}$ is the associated rotlet (or couplet) tensor \citep{pozrikidis2011}. In using the cross product, we follow the convention $[\mathbf{r}\times\mathbf{f}^{\prime}]_i=\epsilon_{ijk}r_jf^\prime_k$ and $[\mathbf{I}\times\mathbf{r}]_{ij}=\epsilon_{jsk}\delta_{is}r_k$, where $\boldsymbol{\epsilon}$ is the third-order permutation tensor. The symmetric and deviatoric first-order force moment (i.e. $\overbracket{\mathbf{f}^{\prime}\mathbf{r}}$) along with the contribution of the {double-layer} potential lead to $\mathbf{S}=\left<\overbracket{\mathbf{f}^{\prime}\mathbf{r}}-2\mu\overbracket{\mathbf{u}^\prime\mathbf{n}}\right>$, namely the stresslet \citep{batchelor1970}. The over-bracket denotes the fully-symmetric and deviatoric part of a tensor which are defined
\begin{align}
\label{sym-trace}
{\overbracket{[~]}}_{ij}&=(1/2)([~]_{ij}+[~]_{ji})-(1/3)[~]_{ss}\delta_{ij}\nonumber,\\
{\overbracket{[~]}}_{ijk}&=({1}/{6})\left([~]_{ijk}+[~]_{ikj}+[~]_{jik}+[~]_{jki}+[~]_{kij}+[~]_{kji}\right)\nonumber\\
&-({1}/{15})\left\{\left([~]_{ssi}+[~]_{sis}+[~]_{iss}\right)\delta_{kj}+\left([~]_{ssj}+[~]_{sjs}+[~]_{jss}\right)\delta_{ik}+\left([~]_{ssk}+[~]_{sks}+[~]_{kss}\right)\delta_{ij}\right\},
\end{align}
for the second and third-order tensors, respectively. 

To determine the flow field decaying as $|\mathbf{x}|^{-3}$, we decompose the third-order tensors $\mathbf{f}^\prime\mathbf{rr}$ and $\mathbf{u}^\prime\mathbf{nr}$ to their irreducible parts (see \citep{andrews1985,andrews1988} for the decomposition technique). Now by taking the second gradient of the Oseen tensor
\begin{align}
\nabla_m\nabla_k J_{ij}&=\frac{\delta_{im}\delta_{jk}+\delta_{jm}\delta_{ik}-\delta_{km}\delta_{ij}}{\left(x_lx_l\right)^{\frac{3}{2}}}+\frac{15x_ix_jx_kx_m}{\left(x_lx_l\right)^{\frac{7}{2}}}
&-\frac{3\left(x_m x_i\delta_{jk}+x_j x_m\delta_{ik}-x_k x_m\delta_{ij}+x_j x_k\delta_{im}+x_i x_k\delta_{jm}+x_i x_j\delta_{km}\right)}{\left(x_lx_l\right)^{\frac{5}{2}}},
\end{align}
we obtain the next order correction for the flow field
\begin{align}
\mathbf{u}^{\prime\text{(3)}}(\mathbf{x})=&-\frac{1}{32\pi\mu}\left[   \left<\overbracket{\mathbf{f}^{\prime}\mathbf{r}\mathbf{r}}-4\mu\overbracket{\mathbf{u}^\prime\mathbf{n}\mathbf{r}}\right>\odot\boldsymbol{\nabla}\mathbf{K}(\mathbf{x}-\mathbf{x}_0)        \right]
-\frac{1}{24\pi\mu}\left[\left<\overbracket{\mathbf{r}(\mathbf{r}\times\mathbf{f}^{\prime})}\right>:\boldsymbol{\nabla}\left[\boldsymbol{\nabla}\times\mathbf{J}(\mathbf{x}-\mathbf{x}_0)\right] \right]\nonumber\\
&-\frac{1}{80\pi\mu}\left[\left<2|\mathbf{r}|^2\mathbf{f}^{\prime}-(\mathbf{r}\cdot\mathbf{f}^{\prime})\mathbf{r} +3\mu\left[4(\mathbf{u}^\prime\cdot\mathbf{n})\mathbf{r} - (\mathbf{u}^\prime\cdot\mathbf{r})\mathbf{n}-(\mathbf{r}\cdot\mathbf{n})\mathbf{u}^\prime\right]\right>\cdot\nabla^2 \mathbf{J}(\mathbf{x}-\mathbf{x}_0)\right].
\end{align}
We may then identify stresslet dipole $\mathbf{S}_\mathcal{D}=\left<\overbracket{\mathbf{f}^{\prime}\mathbf{r}\mathbf{r}}-4\mu\overbracket{\mathbf{u}^\prime\mathbf{n}\mathbf{r}}\right>$, rotlet dipole $\mathbf{C}_\mathcal{D}=\left<\overbracket{\mathbf{r}(\mathbf{r}\times\mathbf{f}^{\prime})}\right>$ and potential dipole $\mathbf{d}=\left<2|\mathbf{r}|^2\mathbf{f}^{\prime}-(\mathbf{r}\cdot\mathbf{f}^{\prime})\mathbf{r} +3\mu\left[4(\mathbf{u}^\prime\cdot\mathbf{n})\mathbf{r} - (\mathbf{u}^\prime\cdot\mathbf{r})\mathbf{n}-(\mathbf{r}\cdot\mathbf{n})\mathbf{u}^\prime\right]\right>$. Finally, we find the flow field around the particle that decays slower than $|\mathbf{x}|^{-4}$ as
\begin{align}
\label{main}
\mathbf{u}^\prime(\mathbf{x})=-\frac{1}{8\pi\mu}\left[\mathbf{F}\cdot\mathbf{J}(\mathbf{x}-\mathbf{x}_0)+\mathbf{L}\cdot\mathbf{C}(\mathbf{x}-\mathbf{x}_0) -\frac{1}{2}\mathbf{S}:\mathbf{K}(\mathbf{x}-\mathbf{x}_0)+\frac{1}{4}\mathbf{S}_\mathcal{D}\odot\boldsymbol{\nabla}\mathbf{K}(\mathbf{x}-\mathbf{x}_0)-\mathbf{C}_\mathcal{D}:\boldsymbol{\Upsilon}(\mathbf{x}-\mathbf{x}_0) +\frac{1}{10}\mathbf{d}\cdot\nabla^2\mathbf{J}(\mathbf{x}-\mathbf{x}_0) \right],
\end{align}
where
\begin{align}
{\Upsilon}_{ijk}= \frac{\epsilon_{iks}x_sx_j+\epsilon_{jks}x_sx_i}{\left(x_lx_l\right)^{\frac{5}{2}}},
\end{align}
is the tensor associated with the rotlet dipole. 

Introducing a more compact notation through 
\begin{align}
\label{strength}
\boldsymbol{\mathsf{S}}&=\left[\mathbf{F},\mathbf{L},\mathbf{S},\mathbf{S}_\mathcal{D},\mathbf{C}_\mathcal{D},\mathbf{d},\cdots\right],\nonumber\\
&=\left[\left<\mathbf{f}^{\prime}\right>,\left<\mathbf{r}\times\mathbf{f}^{\prime}\right>,\left<\overbracket{\mathbf{f}^{\prime}\mathbf{r}}-2\mu\overbracket{\mathbf{u}^\prime\mathbf{n}}\right>,\left<\overbracket{\mathbf{f}^{\prime}\mathbf{r}\mathbf{r}}-4\mu\overbracket{\mathbf{u}^\prime\mathbf{n}\mathbf{r}}\right>,\left<\overbracket{\mathbf{r}(\mathbf{r}\times\mathbf{f}^{\prime})}\right>,\left<2|\mathbf{r}|^2\mathbf{f}^{\prime}-(\mathbf{r}\cdot\mathbf{f}^{\prime})\mathbf{r} +3\mu\left[4(\mathbf{u}^\prime\cdot\mathbf{n})\mathbf{r} - (\mathbf{u}^\prime\cdot\mathbf{r})\mathbf{n}-(\mathbf{r}\cdot\mathbf{n})\mathbf{u}^\prime\right]\right>,\cdots\right],
\end{align}
and
\begin{align}
\boldsymbol{\mathsf{J}}&=\left[\mathbf{J},\mathbf{C},(-1/2)\mathbf{K},(1/4)\boldsymbol{\nabla}\mathbf{K},-\boldsymbol{\Upsilon},(1/10) \nabla^2\mathbf{J}, \cdots\right],
\end{align}
we obtain
\begin{align}
\label{compact}
\mathbf{u}^\prime(\mathbf{x})=-\frac{1}{8\pi\mu}\boldsymbol{\mathsf{S}}\odot\boldsymbol{\mathsf{J}}.
\end{align}
We note in particular that 
\begin{align}
\label{strength2}
\boldsymbol{\mathsf{S}} = \boldsymbol{\mathsf{F}}^{\prime}+\mu\left<\boldsymbol{\mathsf{D}}^{\prime}\right>,
\end{align}
represents the strengths of the multipoles of $\boldsymbol{\mathsf{J}}$ which contains force moments from the single-layer integral,
\begin{align}
{\boldsymbol{\mathsf{F}}}^\prime=\left[\left<\mathbf{f}^{\prime}\right>,\left<\mathbf{r}\times\mathbf{f}^{\prime}\right>,\left<\overbracket{\mathbf{r}\mathbf{f}^{\prime}}\right>,\left<\overbracket{\mathbf{f}^{\prime}\mathbf{r}\mathbf{r}}\right>,\left<\overbracket{\mathbf{r}(\mathbf{r}\times\mathbf{f}^{\prime})}\right>,\left<2|\mathbf{r}|^2\mathbf{f}^{\prime}-(\mathbf{r}\cdot\mathbf{f}^{\prime})\mathbf{r}\right>, \cdots\right],
\end{align}
and terms due to surface disturbance velocity from the double-layer,
\begin{align}
{\boldsymbol{\mathsf{D}}}^{\prime}=\left[\mathbf{0},\mathbf{0},-2\overbracket{\mathbf{n}\mathbf{u}^\prime},-4\overbracket{\mathbf{u}^\prime\mathbf{n}\mathbf{r}},\mathbf{0},12\left(\mathbf{u}^\prime\cdot\mathbf{n}\right)\mathbf{r} - 3\left(\mathbf{u}^\prime\cdot\mathbf{r}\right)\mathbf{n}-3(\mathbf{r}\cdot\mathbf{n})\mathbf{u}^\prime, \cdots\right].
\end{align}
Finally, we note $\boldsymbol{\mathsf{S}}=\boldsymbol{\mathsf{S}}^\prime$ because all terms in $\boldsymbol{\mathsf{S}}^\infty$ are zero as the boundary integral equation vanishes identically in that case \cite{pozrikidis1992}.

To summarize, the second-order force moment is decomposed into a rotlet dipole, a stresslet dipole and a potential dipole, with contributions from the {double-layer} potentials. To better understand the physical interpretation of these force moments, consider an active particle that propels itself forward using flagella (e.g., \textit{E. coli}). The force exerted by the flagella is completely balanced by the drag force on the body and since the distribution of these two forces is separated in position (e.g., tail and head), the induced flow field is captured by a stresslet. However, the length scales of the cell body and the flagella differ, often by orders of magnitude, and so this asymmetry gives rise to a stresslet dipole. For a case wherein flagella use rotation to generate thrust, the cell body must counter-rotate to maintain the torque-free motion thereby generating a rotlet dipole. Finally, the finite size of the cell body accounts for the presence of a potential dipole (further discussion may be found in Refs. \cite{spagnolie12,smith2009}).

We should emphasize that the multipole expansion given in \eqref{compact} is valid for an active (or passive) particle of any arbitrary shape, in a viscous fluid. Given that $\boldsymbol{\mathsf{J}}$ is generic for any unbounded single particle, determining the flow field is reduced to finding the strengths $\boldsymbol{\mathsf{S}}$. Thus, both traction $\mathbf{f}^{\prime}$ and disturbance surface velocity $\mathbf{u}^\prime\left(\mathbf{x}\in\partial\mathcal{B}\right)$ are needed. Although the latter may be explicitly prescribed (e.g. through slip velocity), finding the traction generally requires solving the flow field in full. However, one can avoid such calculations by using the Lorentz reciprocal theorem for the Stokes flow \citep{stone1996,elfring2015,lauga2016} to calculate \textit{moments} of $\mathbf{f}^{\prime}$. In the following, we employ the general framework given in \citep{elfring2017} to find the force moments and hence the multipole strengths $\boldsymbol{\mathsf{S}}$. Recovering the expressions for force, torque and stresslet, we report explicit formulas for the stresslet dipole, rotlet dipole and potential dipole for an arbitrarily-shaped active particle.
\section{Evaluating the force moments of an active particle}
\label{reciprocal}
We are interested in the motion of an active particle with boundary conditions
\begin{align}
\mathbf{u}\left(\mathbf{x}\in\partial\mathcal{B}\right)=\mathbf{U} + \boldsymbol{\Omega}\times\mathbf{r}+\mathbf{u}^s,
\end{align}
where $\mathbf{U}$ and $\boldsymbol{\Omega}$ are the rigid-body translation and rotation of the particle while $\mathbf{u}^s$ is a velocity due to surface activity, such as diffusiophoretic slip or a swimming gait \cite{michelin14,elfring15}. As a dual (or auxiliary) problem, here denoted by a hat, we take the passive motion of a rigid body of the same instantaneous shape,
\begin{align}
\hat{\mathbf{u}}\left(\mathbf{x}\in\partial\mathcal{B}\right)=\hat{\mathbf{U}} + \hat{\boldsymbol{\Omega}}\times\mathbf{r}.
\end{align}

The reciprocal theorem indicates that the virtual power of the motion of these two bodies is equal \citep{happel1981}, namely
\begin{align}
\label{recip}
\hat{\mu}\left<\mathbf{n}\cdot{\boldsymbol{\sigma}}^{\prime}\cdot\hat{\mathbf{u}}^{\prime}\right>=\mu\left<\mathbf{n}\cdot\hat{\boldsymbol{\sigma}}^{\prime}\cdot{\mathbf{u}}^{\prime}\right>,
\end{align}
where $\hat{\mathbf{u}}^{\prime}=\hat{\mathbf{u}}-\hat{\mathbf{u}}^{\infty}$ and $\hat{\boldsymbol{\sigma}}^{\prime}=\hat{\boldsymbol{\sigma}}-\hat{\boldsymbol{\sigma}}^{\infty}$ are the disturbance flow and stress field for the auxiliary problem, respectively. As we will show below, by using (operators of) the dual problem, the force moments of the active particle may be obtained without resolution of the disturbance field $\mathbf{u}^\prime$ nor traction $\mathbf{f}^\prime$. 

Following \citet{elfring2017}, we expand the background flow of the auxiliary problem around a point in the body $\mathbf{x}_0$ as
\begin{align}
\hat{\mathbf{u}}^\infty(\mathbf{x}\in\partial\mathcal{B})= \hat{\mathbf{U}}^\infty\left(\mathbf{x}_0\right)+\mathbf{r}\cdot\boldsymbol{\nabla}\hat{\mathbf{u}}^\infty\left(\mathbf{x}_0\right)+\frac{1}{2}\mathbf{r}\mathbf{r}:\boldsymbol{\nabla}\boldsymbol{\nabla}\hat{\mathbf{u}}^\infty\left(\mathbf{x}_0\right)+\cdots,
\end{align}
which by decomposing $\boldsymbol{\nabla}\hat{\mathbf{u}}^\infty$ and $\boldsymbol{\nabla}\boldsymbol{\nabla}\hat{\mathbf{u}}^\infty$ to their irreducible parts, can be rewritten
\begin{align}
\label{aux}
\hat{\mathbf{u}}^\infty(\mathbf{x}\in\partial\mathcal{B})= \hat{\mathbf{U}}^\infty+\hat{\boldsymbol{\Omega}}^\infty\times\mathbf{r}+\mathbf{r}\cdot\hat{\mathbf{E}}^\infty+\mathbf{r}\mathbf{r}:\hat{\boldsymbol{\Gamma}}^\infty+\left(\boldsymbol{\epsilon}\cdot\mathbf{r}\right)\mathbf{r}:\hat{\boldsymbol{\Lambda}}^\infty+\left(2|\mathbf{r}|^2\mathbf{I}-\mathbf{r}\mathbf{r}\right)\cdot\hat{\mathbf{e}}^\infty+\cdots.
\end{align}
Here, $ \hat{\mathbf{U}}^\infty$ and $\hat{\boldsymbol{\Omega}}^\infty$ indicate the translation and rotation of the background flow of auxiliary problem at $\mathbf{x}_0$. $\hat{\mathbf{E}}^\infty=\overbracket{\boldsymbol{\nabla}\hat{\mathbf{u}}^\infty}$ and $\hat{\boldsymbol{\Gamma}}^\infty=(1/2)\overbracket{\boldsymbol{\nabla}\boldsymbol{\nabla}\hat{\mathbf{u}}^\infty}$ are the fully symmetric and deviatoric ($\hat{E}_{ii}^\infty=0$; $\hat{{\Gamma}}_{iij}^\infty=\hat{{\Gamma}}_{iji}^\infty=\hat{{\Gamma}}_{jii}^\infty=0$) second and third-order tensors, respectively. $\hat{\boldsymbol{\Lambda}}^\infty=\overbracket{\boldsymbol{\nabla}\left(\boldsymbol{\nabla}\times\hat{\mathbf{u}}^\infty\right)}$ is a second-order symmetric tensor and $\hat{\mathbf{e}}^\infty=(1/10)\nabla^2\hat{\mathbf{u}}^\infty$. Using this expansion and relying on the linearity of the Stokes equations, we can write
\begin{align}
\hat{p}^{\prime}&=\hat{\mu}\hat{\boldsymbol{\mathsf{P}}}\odot\hat{\boldsymbol{\mathsf{U}}}^\prime,\\
\hat{\mathbf{u}}^{\prime}&=\hat{\boldsymbol{\mathsf{G}}}\odot\hat{\boldsymbol{\mathsf{U}}}^\prime,\\
\hat{\boldsymbol{\sigma}}^{\prime}&=\hat{\mu}\hat{\boldsymbol{\mathsf{T}}}\odot\hat{\boldsymbol{\mathsf{U}}}^{\prime},\\
\hat{{\boldsymbol{\mathsf{F}}}}^\prime&=-\hat{{\boldsymbol{\mathsf{R}}}}\odot\hat{{\boldsymbol{\mathsf{U}}}}^\prime,
\end{align}
wherein the velocity gradients
\begin{align}
\hat{\boldsymbol{\mathsf{U}}}^{\prime}&=\left[\hat{\mathbf{U}}-\hat{\mathbf{U}}^\infty,\hat{\boldsymbol{\Omega}}-\hat{\boldsymbol{\Omega}}^\infty,-\hat{\mathbf{E}}^\infty,-\hat{\boldsymbol{\Gamma}}^\infty,-\hat{\boldsymbol{\Lambda}}^\infty,-\hat{\mathbf{e}}^\infty,\cdots\right],
\end{align}
are linearly mapped to the disturbance pressure, velocity, and stress fields by
\begin{align}
\hat{\boldsymbol{\mathsf{P}}}&=\left[\hat{\mathbf{P}}_U,\hat{\mathbf{P}}_\Omega,\hat{\mathbf{P}}_E,\hat{\mathbf{P}}_{\Gamma},\hat{\mathbf{P}}_{\Lambda},\hat{\mathbf{P}}_e,\cdots\right],\\
\hat{\boldsymbol{\mathsf{G}}}&=\left[\hat{\mathbf{G}}_U,\hat{\mathbf{G}}_\Omega,\hat{\mathbf{G}}_E,\hat{\mathbf{G}}_{\Gamma},\hat{\mathbf{G}}_{\Lambda},\hat{\mathbf{G}}_e,\cdots\right],\\
\hat{\boldsymbol{\mathsf{T}}}&=\left[\hat{\mathbf{T}}_U,\hat{\mathbf{T}}_\Omega,\hat{\mathbf{T}}_E,\hat{\mathbf{T}}_{\Gamma},\hat{\mathbf{T}}_{\Lambda},\hat{\mathbf{T}}_e,\cdots\right],
\end{align}
which are functions of the position in space and the geometry of the particle and each term maintains the symmetry of the term, against which it operates. $\hat{{\boldsymbol{\mathsf{R}}}}$ is the grand resistance tensor that linearly maps velocity moments to force moments in the dual problem. The symmetry of the stress tensor implies that all components of $\hat{\boldsymbol{\mathsf{T}}}$ are symmetric in their first two (or non-contracted) indices. Under these definitions, the reciprocal theorem can be rewritten as
\begin{align}
\label{recip2}
\left<\left(\mathbf{n}\cdot{\boldsymbol{\sigma}}^{\prime}\right)\cdot\hat{\boldsymbol{\mathsf{G}}}\right>\odot\hat{\boldsymbol{\mathsf{U}}}^\prime=\mu\left<\mathbf{u}^\prime\cdot\left(\mathbf{n}\cdot\hat{\boldsymbol{\mathsf{T}}}\right)\right>\odot\hat{\boldsymbol{\mathsf{U}}}^\prime.
\end{align}
Importantly, given the boundary condition of the dual problem and the decomposition of the background field, we know the terms in the operator $\hat{\boldsymbol{\mathsf{G}}}(\mathbf{x}\in\partial\mathcal{B})$ on the boundary of the particle, which may be defined as
\begin{align}
\label{U}
\hat{G}_{U,ij}&=\delta_{ij},\\
\hat{G}_{\Omega,ij}&=\epsilon_{ijs}r_s,\\
\hat{G}_{E,ijk}&=\overbracket{\delta_{ij}r_k}^{jk},\\
\hat{G}_{\Gamma,ijkm}&=\overbracket{\delta_{ij}r_kr_m}^{jkm},\\
\hat{G}_{\Lambda,ijk}&=\overbracket{\epsilon_{ijs}r_sr_k}^{jk},\\
\label{e}
\hat{G}_{e,ijkm}&=2\delta_{ij}r_sr_s-r_ir_j,\\
 &\ \ \vdots\nonumber
\end{align}
where the over-brackets are identical to \eqref{sym-trace} but only operate over the specified indices. In this way, the operator acts precisely to map the traction $\mathbf{f}'=\mathbf{n}\cdot{\boldsymbol{\sigma}}^{\prime}$ to the force moments
\begin{align}
\label{force-moments}
{\boldsymbol{\mathsf{F}}}^\prime=\left<\mathbf{f}'\cdot\hat{\boldsymbol{\mathsf{G}}}\right>=\left[\left<\mathbf{f}^{\prime}\right>,\left<\mathbf{r}\times\mathbf{f}^{\prime}\right>,\left<\overbracket{\mathbf{r}\mathbf{f}^{\prime}}\right>,\left<\overbracket{\mathbf{f}^{\prime}\mathbf{r}\mathbf{r}}\right>,\left<\overbracket{\mathbf{r}(\mathbf{r}\times\mathbf{f}^{\prime})}\right>,\left<2|\mathbf{r}|^2\mathbf{f}^{\prime}-(\mathbf{r}\cdot\mathbf{f}^{\prime})\mathbf{r}\right>, \cdots\right].
\end{align}

Now, given that $\hat{\boldsymbol{\mathsf{U}}}^\prime$ is arbitrarily chosen, we may discard it from both sides of equation \eqref{recip2}. Applying the boundary conditions on $\mathbf{u}^{\prime}$, and also expressing the rigid body translation and rotation as $\boldsymbol{\mathsf{U}}=[\mathbf{U},\boldsymbol{\Omega},\mathbf{0},\cdots]$, equation \eqref{recip2} can be reduced to \citep{elfring2017}
\begin{align}
\label{gwynn-eq}
{\boldsymbol{\mathsf{F}}}^\prime&=-\frac{\mu}{\hat{\mu}}\hat{{\boldsymbol{\mathsf{R}}}}\odot\boldsymbol{\mathsf{U}}+\mu\left<\left(\mathbf{u}^s-\mathbf{u}^\infty\right)\cdot\left(\mathbf{n}\cdot\hat{\boldsymbol{\mathsf{T}}}\right)\right>.
\end{align}

Then by way of \eqref{strength2}, one obtains the multipole strengths as
\begin{align}
\label{babak-eq}
\boldsymbol{\mathsf{S}}=-\frac{\mu}{\hat{\mu}}\hat{{\boldsymbol{\mathsf{R}}}}\odot\boldsymbol{\mathsf{U}}+\mu\left<\left(\mathbf{u}^s-\mathbf{u}^\infty\right)\cdot\left(\mathbf{n}\cdot\hat{\boldsymbol{\mathsf{T}}}\right)+{\boldsymbol{\mathsf{D}}}^\prime\right>,
\end{align} 
where the double-layer potential contribution,
\begin{align}
{\boldsymbol{\mathsf{D}}}^\prime=\left[\mathbf{0},\mathbf{0},-2\overbracket{\mathbf{n}\left(\mathbf{u}^s-\mathbf{u}^\infty\right)},-4\overbracket{\left(\mathbf{u}^s-\mathbf{u}^\infty\right)\mathbf{n}\mathbf{r}},\mathbf{0},12\left[\left(\mathbf{u}^s-\mathbf{u}^\infty\right)\cdot\mathbf{n}\right]\mathbf{r} - 3\left[\left(\mathbf{u}^s-\mathbf{u}^\infty\right)\cdot\mathbf{r}\right]\mathbf{n}-3(\mathbf{r}\cdot\mathbf{n})\left(\mathbf{u}^s-\mathbf{u}^\infty\right), \cdots\right],
\end{align}
is simplified as the terms associated with rigid-body motion integrate to zero \citep{pozrikidis1992}. 

Equation \eqref{babak-eq} provides the tensorial relationship between the boundary motion and the strength of multipoles for any arbitrarily-shaped active particle in Stokes flow. We note that the multipole strengths are split into terms arising from the rigid-body motion of the particle, $\boldsymbol{\mathsf{U}}$, and those associated with the (disturbance) surface activity $\mathbf{u}^s-\mathbf{u}^\infty$. 
Using this equation, one can derive explicit formulas for $\boldsymbol{\mathsf{S}}$ provided $\hat{\boldsymbol{\mathsf{T}}}$ is known, as we illustrate in the following.

We begin with $\mathbf{F}$ and $\mathbf{L}$. In self-propulsion, in the absence of any external force and torque, the net force and torque on the particle are strictly zero. However, to find the net translational and rotational velocity (which are unknown at this point), we may use the reciprocal theorem for the force and torque. Upon setting $\mathbf{F}=\mathbf{0}$ and $\mathbf{L}=\mathbf{0}$ in equation \eqref{babak-eq} we may solve for $\boldsymbol{\mathsf{U}}$ directly
\begin{align}
\label{resistance}
\left[ {\begin{array}{cc}
\mathbf{U} \\
\boldsymbol{\Omega} \\
  \end{array} } \right]
  =\frac{\hat{\mu}}{{\mu}}
  \left({\begin{array}{cc}
   \hat{\mathbf{R}}_{FU} & \hat{\mathbf{R}}_{LU} \\
   \hat{\mathbf{R}}_{F\Omega} & \hat{\mathbf{R}}_{L\Omega} \\
  \end{array} } \right)
  ^{-1}\cdot  \left[ {\begin{array}{cc}
{\mathbf{F}^s} \\
  { \mathbf{L}^s }\\
  \end{array} } \right],
\end{align}
where $\hat{\mathbf{R}}_{FU}=-\hat{\mu}\left<{\mathbf{n}}\cdot\hat{\mathbf{T}}_{U}\right>$, $\hat{\mathbf{R}}_{F\Omega}=-\hat{\mu}\left<{\mathbf{n}}\cdot\hat{\mathbf{T}}_{\Omega}\right>$, $\hat{\mathbf{R}}_{LU}=-\hat{\mu}\left<\mathbf{r}\times\left({\mathbf{n}}\cdot\hat{\mathbf{T}}_{U}\right)\right>$ and $\hat{\mathbf{R}}_{L\Omega}=-\hat{\mu}\left<\mathbf{r}\times\left({\mathbf{n}}\cdot\hat{\mathbf{T}}_{\Omega}\right)\right>$ are the components of the grand resistance tensor $\hat{\boldsymbol{\mathsf{R}}}$ associated with rigid-body motion. Here ${\mathbf{F}^s}=\mu\left<\left(\mathbf{u}^s-\mathbf{u}^{\infty}\right)\cdot\left(\mathbf{n}\cdot\hat{\mathbf{T}}_U\right)\right>$ is the hydrodynamic force arising solely from the surface activities (often referred to as the \textit{thrust}) and ${\mathbf{L}^s}=\mu\left<\left(\mathbf{u}^s-\mathbf{u}^{\infty}\right)\cdot\left(\mathbf{n}\cdot\hat{\mathbf{T}}_\Omega\right)\right>$ is surface activity driven torque. Equation \eqref{resistance}  simply illustrates the balance between the force and torque generated by the surface activities and the hydrodynamic drag.

We may now determine other components of $\boldsymbol{\mathsf{S}}$ by using \eqref{babak-eq} at higher orders. We obtain
\begin{align}
\label{stresslet1}
\mathbf{S}&=-\frac{\mu}{\hat{\mu}}\left(\hat{\mathbf{R}}_{SU}\cdot\mathbf{U}+\hat{\mathbf{R}}_{S\Omega}\cdot\boldsymbol{\Omega}\right)+{\mathbf{S}^s},\\
\mathbf{S}_\mathcal{D}&=-\frac{\mu}{\hat{\mu}}\left(\hat{\mathbf{R}}_{S_\mathcal{D}U}\cdot\mathbf{U} +\hat{\mathbf{R}}_{S_\mathcal{D}\Omega}\cdot\boldsymbol{\Omega}\right)+{\mathbf{S}_\mathcal{D}^s}, \\
\mathbf{C}_\mathcal{D}&=-\frac{\mu}{\hat{\mu}}\left(\hat{\mathbf{R}}_{C_\mathcal{D}U}\cdot\mathbf{U} +\hat{\mathbf{R}}_{C_\mathcal{D}\Omega}\cdot\boldsymbol{\Omega}\right)+{\mathbf{C}_\mathcal{D}^s}, \\
\mathbf{d}&= -\frac{\mu}{\hat{\mu}}\left(\hat{\mathbf{R}}_{dU}\cdot\mathbf{U} +\hat{\mathbf{R}}_{d\Omega}\cdot\boldsymbol{\Omega}\right)+{\mathbf{d}^s},
\end{align}
where the resistance tensors may be written in terms of $\hat{\boldsymbol{\mathsf{T}}}$ as follows
\begin{align}
\label{rsu}
{\hat{R}}_{SU,ijk} &=-\hat{\mu}\left<\overbracket{n_s\hat{T}_{U,sik}r_j}^{ij}\right>,\\
{\hat{R}}_{S\Omega,ijk} &=-\hat{\mu}\left<\overbracket{n_s\hat{T}_{\Omega,sik}r_j}^{ij}\right>,\\
\hat{{R}}_{S_\mathcal{D}U,ijkm}&=-\hat{\mu}\left<\overbracket{n_s\hat{T}_{{U},sim}r_j r_k}^{ijk}\right>,\\
\hat{{R}}_{S_\mathcal{D}\Omega,ijkm}&=-\hat{\mu}\left<\overbracket{n_s\hat{T}_{\Omega,sim}r_j r_k}^{ijk}\right>,\\
\hat{{R}}_{C_\mathcal{D}U,ijk}&=-\hat{\mu}\left<\overbracket{n_s\hat{T}_{U,snk}\epsilon_{jmn}r_ir_m}^{ij}\right>,\\
\label{rcdu}
\hat{{R}}_{C_\mathcal{D}\Omega,ijk}&=-\hat{\mu}\left<\overbracket{n_s\hat{T}_{\Omega,snk}\epsilon_{jmn}r_ir_m}^{ij}\right>,\\
\hat{{R}}_{dU,ij}&=-\hat{\mu}\left<2n_s\hat{T}_{U,sij}r_l r_l -n_s\hat{T}_{U,smj}r_mr_i\right>,\\
\hat{{R}}_{d\Omega,ij}&=-\hat{\mu}\left<2n_s\hat{T}_{\Omega,sij}r_l r_l -n_s\hat{T}_{\Omega,smj}r_mr_i\right>,
\end{align}
and the contributions of the surface activities are likewise
\begin{align}
{\mathbf{S}^s}&=\mu\left<\overbracket{\left(\mathbf{u}^s-\mathbf{u}^\infty\right)\mathbf{n}:\hat{\mathbf{T}}_E}-2\overbracket{\left(\mathbf{u}^s-\mathbf{u}^\infty\right)\mathbf{n}}\right>,\\
{\mathbf{S}_\mathcal{D}^s}&=\mu\left<\overbracket{\left(\mathbf{u}^s-\mathbf{u}^\infty\right)\mathbf{n}:\hat{\mathbf{T}}_{\Gamma}}-4\overbracket{\left(\mathbf{u}^s-\mathbf{u}^\infty\right)\mathbf{n}\mathbf{r}}\right>,\\
{\mathbf{C}_\mathcal{D}^s}&=\mu\left<\overbracket{\left(\mathbf{u}^s-\mathbf{u}^\infty\right)\mathbf{n}:\hat{\mathbf{T}}_{\Lambda}}\right>,\\
{\mathbf{d}^s}&=\mu\left<\left(\mathbf{u}^s-\mathbf{u}^\infty\right)\mathbf{n}:\hat{\mathbf{T}}_e+12\left[\mathbf{n}\cdot\left(\mathbf{u}^s-\mathbf{u}^\infty\right)\right]\mathbf{r}-3\left[\mathbf{r}\cdot\left(\mathbf{u}^s-\mathbf{u}^\infty\right)\right]\mathbf{n}-3\left(\mathbf{r}\cdot\mathbf{n}\right)\left(\mathbf{u}^s-\mathbf{u}^\infty\right)\right>.
\end{align} 
We should emphasize that all components of $\hat{\boldsymbol{\mathsf{T}}}$ are unique for a given particle geometry. Therefore by finding them once, we can determine the force moments for any prescribed surface activity provided the shape does not change.

\subsection{Sphere}
We now resolve the force moments of an active spherical particle, using the expressions reported in the previous section. We take $\mathbf{x}_0$ to be the center of the sphere $\mathbf{r}=a\mathbf{n}$, where $a$ is the radius. Details of the auxiliary flow field and stress field corresponding to each force moment (i.e., $\hat{\boldsymbol{\mathsf{P}}}$, $\hat{\boldsymbol{\mathsf{G}}}$ and $\hat{\boldsymbol{\mathsf{T}}}$) are reported in the appendix.

Having $\hat{\mathbf{T}}_U$ and $\hat{\mathbf{T}}_\Omega$ at hand, the rigid-body resistance tensors can be evaluated. We find $\hat{\mathbf{R}}_{FU}=6\pi\hat{\mu} a\mathbf{I}$, $\hat{\mathbf{R}}_{L\Omega}=8\pi\hat{\mu} a^3\mathbf{I}$, $\hat{\mathbf{R}}_{dU}=10\pi\hat{\mu} a^3\mathbf{I}$ and $\left[\hat{\mathbf{R}}_{F\Omega},\hat{\mathbf{R}}_{LU},\hat{\mathbf{R}}_{SU},\hat{\mathbf{R}}_{S\Omega}, \hat{\mathbf{R}}_{S_\mathcal{D}U},\hat{\mathbf{R}}_{S_\mathcal{D}\Omega}, \hat{\mathbf{R}}_{C_\mathcal{D}U},\hat{\mathbf{R}}_{C_\mathcal{D}\Omega},\hat{\mathbf{R}}_{d\Omega}\right]=\mathbf{0}$. 

The force and torque are respectively
\begin{align}
\mathbf{F} &= -6\pi a\mu \mathbf{U}-\frac{3\mu}{2a}\left<\mathbf{u}^s-\mathbf{u}^\infty\right> = -6\pi a\mu \big(\mathbf{U}-(1+\frac{a^2}{6}\nabla^2)\mathbf{U}^\infty\big)-\frac{3\mu}{2a}\left<\mathbf{u}^s\right>,\\
\mathbf{L} &= -8\pi a^3\mu\mathbf{\Omega}+3\mu\left<(\mathbf{u}^s-\mathbf{u}^\infty)\times\mathbf{n}\right>=-8\pi a^3\mu\left(\mathbf{\Omega}-\mathbf{\Omega}^\infty\right)+3\mu\left<\mathbf{u}^s\times\mathbf{n}\right>,
\end{align}
where $\mathbf{U}^\infty$ and $\mathbf{\Omega}^\infty$ are the velocity and rotation rate of the background flow at the center of the sphere. If the particle is passive, $\mathbf{u}^s=\mathbf{0}$, we recover Fax\'en's first and second laws as expected. In the absence of an external force and torque, the rigid-body translation and rotation of spherical active particle with surface velocities $\mathbf{u}^s$ are given by
\begin{align}
\mathbf{U} &= -\frac{1}{4\pi a^2}\left<\mathbf{u}^s-\mathbf{u}^\infty\right>,\\
\mathbf{\Omega}&= -\frac{3}{8\pi a^4}\left<\mathbf{r}\times\left(\mathbf{u}^s-\mathbf{u}^\infty\right)\right>,
\end{align}
as first shown by \citet{anderson91} and later generalized \cite{stone1996,elfring2015}.  
Using the expression for stresslet given in equation \eqref{stresslet1}, we find 
\begin{align}
\label{stresslet}
\mathbf{S}&=-5\mu\left<\overbracket{(\mathbf{u}^s-\mathbf{u}^\infty)\mathbf{n}}\right>-\frac{2}{3}\mu\left<(\mathbf{u}^s-\mathbf{u}^\infty)\cdot\mathbf{n}\right>\mathbf{I},\nonumber\\
&=\frac{20\pi\mu a^3 }{3}\left(1+\frac{a^2}{10}\nabla^2\right)\mathbf{E}^\infty-5\mu\left<\overbracket{\mathbf{u}^{s}\mathbf{n}}\right>-\frac{2}{3}\mu\left<\mathbf{u}^{s}\cdot\mathbf{n}\right>\mathbf{I},
\end{align}
where $\mathbf{E}^\infty=\overbracket{\boldsymbol{\nabla}\mathbf{u}^\infty\left(\mathbf{x}_0\right)}$. We note that this expression for the stresslet amends a typographical error in the results of \citet{lauga2016} (equation (10) in their reference). When the sphere is passive, equation \eqref{stresslet} recovers Fax\'en's third law.
By using $\mathbf{\hat{{T}}}_{\Gamma}$, we determine the stresslet dipole 
\begin{align}
\mathbf{S}_\mathcal{D}&=-\frac{35}{4}\mu a\left<\overbracket{\left(\mathbf{u}^{s}-\mathbf{u}^\infty\right)\mathbf{n}\mathbf{n}}\right>,\nonumber\\
&=\frac{14\mu\pi a^5}{3}\left(1+\frac{a^2}{14}\nabla^2\right)\boldsymbol\Gamma^\infty-\frac{35}{4}\mu a\left<\overbracket{\mathbf{u}^{s}\mathbf{n}\mathbf{n}}\right>,
\end{align}
where $\boldsymbol\Gamma^\infty=\left({1}/{2}\right)\overbracket{\boldsymbol{\nabla}\boldsymbol{\nabla}\mathbf{u}^\infty\left(\mathbf{x}_0\right)}$. 
The rotlet dipole is then similarly found
\begin{align}
\mathbf{C}_\mathcal{D}&=4\mu a \left<\overbracket{\left[\left(\mathbf{u}^{s}-\mathbf{u}^\infty\right)\times\mathbf{n}\right]\mathbf{n}}\right>,\nonumber\\
&=\frac{16\mu\pi a^5}{15}\boldsymbol{\Lambda}^\infty+4\mu a \left<\overbracket{\left[\mathbf{u}^{s}\times\mathbf{n}\right]\mathbf{n}}\right>,
\end{align}
where $\boldsymbol{\Lambda}^\infty=\overbracket{\boldsymbol{\nabla}\left(\boldsymbol{\nabla}\times\mathbf{u}^\infty\right)\left(\mathbf{x}_0\right)}$. 
Finally, for the potential dipole, we arrive at
\begin{align}
\mathbf{d}&=-10\pi \mu a^3\mathbf{U}+\frac{15a\mu}{2}\left<2\left[\left(\mathbf{u}^{s}-\mathbf{u}^\infty\right)\cdot\mathbf{n}\right]\mathbf{n}-\mathbf{u}^{s}+\mathbf{u}^\infty\right>,\nonumber\\
&=-10\pi \mu a^3\left(\mathbf{U}-\mathbf{U}^\infty\right)+30\mu\pi a^5 \mathbf{d}^\infty+\frac{15a\mu}{2}\left<2\left(\mathbf{u}^{s}\cdot\mathbf{n}\right)\mathbf{n}-\mathbf{u}^{s}\right>,
\end{align}
with $\mathbf{d}^\infty=\left(1/10\right)\nabla^2\mathbf{u}^\infty\left(\mathbf{x}_0\right)$. In total, for a spherical active particle, we have
\begin{align}
\boldsymbol{\mathsf{S}}=\bigg[&-6\pi a\mu \mathbf{U}-\frac{3\mu}{2a}\left<\mathbf{u}^s-\mathbf{u}^\infty\right>,-8\pi a^3\mu\mathbf{\Omega}+3\mu\left<(\mathbf{u}^s-\mathbf{u}^\infty)\times\mathbf{n}\right>,-5\mu\left<\overbracket{\left(\mathbf{u}^{s}-\mathbf{u}^\infty\right)\mathbf{n}}\right>-\frac{2}{3}\mu\left<\left(\mathbf{u}^{s}-\mathbf{u}^\infty\right)\cdot\mathbf{n}\right>\mathbf{I},\nonumber\\
&-\frac{35}{4}\mu a\left<\overbracket{\left(\mathbf{u}^{s}-\mathbf{u}^\infty\right)\mathbf{n}\mathbf{n}}\right>,4\mu a \left<\overbracket{\left[\left(\mathbf{u}^{s}-\mathbf{u}^\infty\right)\times\mathbf{n}\right]\mathbf{n}}\right>,-10\pi \mu a^3\mathbf{U}+\frac{15a\mu}{2}\left<2\left[\left(\mathbf{u}^{s}-\mathbf{u}^\infty\right)\cdot\mathbf{n}\right]\mathbf{n}-\mathbf{u}^{s}+\mathbf{u}^\infty\right>,\cdots\bigg].
\end{align}
\subsection{Generalized squirmer}
\begin{figure}[H]
\begin{center}
\includegraphics[scale=0.5]{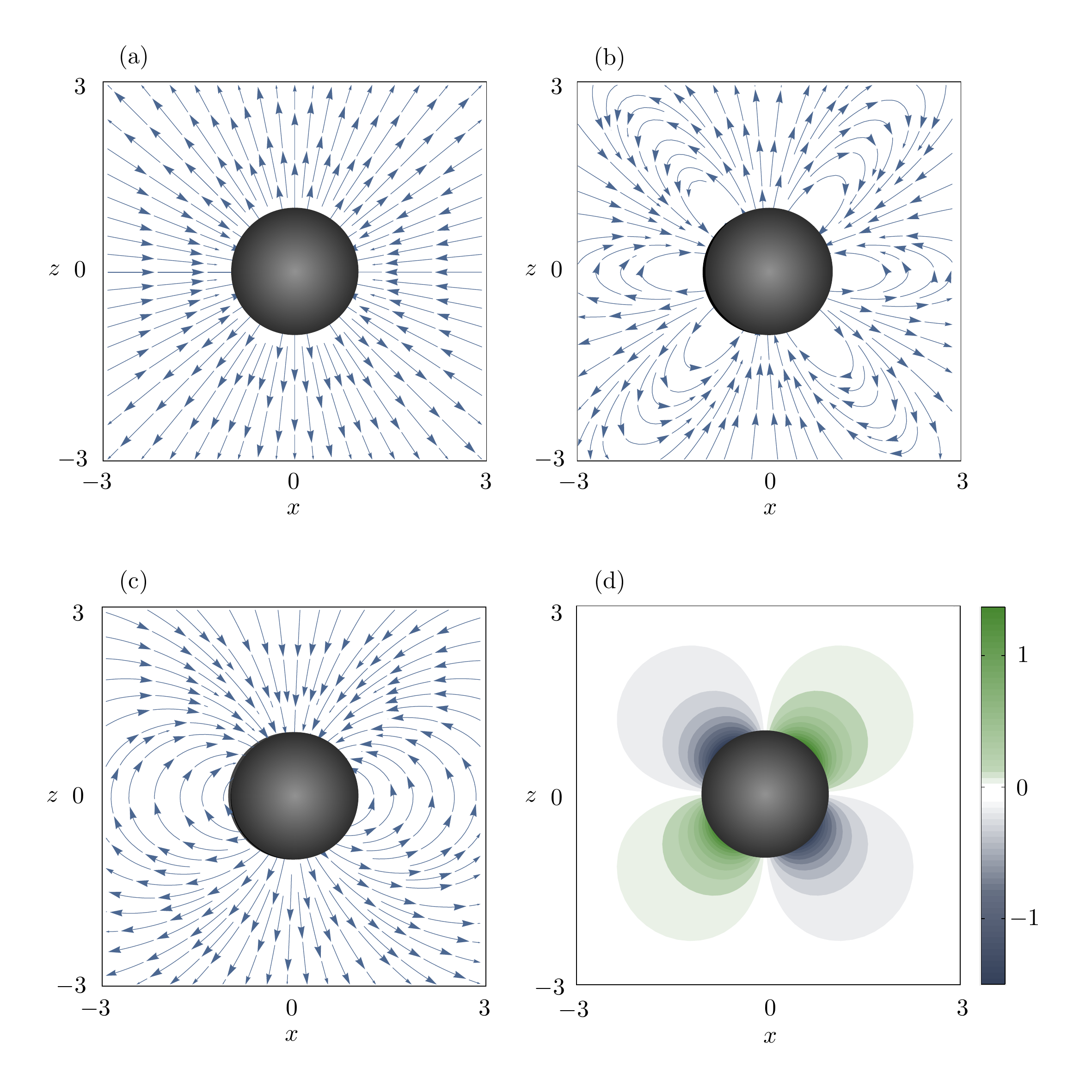}
\caption{Flow fields induced by non-zero force moments of an axisymmetric squirmer of radius 1, using expressions given in \eqref{stresslet-pak} to \eqref{potdipole-pak}: (a) Flow field due to a stresslet, for which we set $B_{02}=1$ and other coefficients to zero. (b) Flow field due to a stresslet dipole with $B_{03}=1$. (c) Flow field induced by a potential dipole with $B_{01}=1$. (d) Flow field due to a rotlet dipole with $C_{02}=1$. In (d), the color density indicates the magnitude of the velocity where positive (negative) values indicate flow into the plane (out of the plane).}
\label{flow-field}
\end{center}
\end{figure}

We now examine the expressions obtained above for the specific case of a sphere with purely tangential surface activity, i.e., a squirmer \citep{lighthill1952,blake1971,pak2014}. One may then express $\mathbf{u}^s={u}^s_\theta\mathbf{e}_\theta+{u}^s_\phi\mathbf{e}_\phi$ in spherical coordinates $(r,\theta,\phi)$ as \citep{pak2014}
\begin{align}
\label{pak1}
u^s_\theta&=\sum_{n=1}^{\infty}\sum_{m=0}^{n}\left[-\frac{2\sin\theta {P_{n}^{m}}^\prime(\xi)}{na^{n+2}}\left(B_{mn}\cos m\phi + \tilde{B}_{mn}\sin m\phi\right)+\frac{m {P_{n}^{m}(\xi)}}{a^{n+1}\sin\theta}\left(\tilde{C}_{mn}\cos m\phi - {C}_{mn}\sin m\phi\right)\right],\\
\label{pak2}
u^s_\phi&=\sum_{n=1}^{\infty}\sum_{m=0}^{n}\left[\frac{\sin\theta {P_{n}^{m}}^\prime(\xi)}{a^{n+1}}\left(C_{mn}\cos m\phi + \tilde{C}_{mn}\sin m\phi\right)+\frac{2m {P_{n}^{m}(\xi)}}{n a^{n+2}\sin\theta}\left(\tilde{B}_{mn}\cos m\phi - {B}_{mn}\sin m\phi\right)\right],
\end{align}
where $P_n^m(\xi)$ is a Legendre function of order $m$ and degree $n$, and the prime in ${P_n^m}^\prime(\xi)$ indicates differentiation with respect to $\xi=\cos\theta$. Here, $B_{mn},\tilde{B}_{mn},C_{mn}$ and $\tilde{C}_{mn}$ are constant coefficients representing different modes of the surface activity. We find the net translational and rotational velocities in terms of these coefficients and Cartesian unit vectors $\mathbf{e}_x, \mathbf{e}_y$ and $\mathbf{e}_z$ as
\begin{align}
\label{U-pak}
\mathbf{U}&=-\frac{4}{3a^3}\left(B_{01}\mathbf{e}_z -B_{11}\mathbf{e}_x-\tilde{B}_{11}\mathbf{e}_y\right),\\
\boldsymbol{\Omega}&=-\frac{1}{a^3}\left(C_{01}\mathbf{e}_z -C_{11}\mathbf{e}_x-\tilde{C}_{11}\mathbf{e}_y\right).
\end{align}
Stresslet, stresslet dipole, rotlet dipole and potential dipole can be similarly determined
\begin{align}
\label{stresslet-pak}
\mathbf{S}=&-\frac{12\pi\mu}{a^2}\left[B_{02}\overbracket{\mathbf{e}_z\mathbf{e}_z}-2B_{12}\overbracket{\mathbf{e}_x\mathbf{e}_z}-2\tilde{B}_{12}\overbracket{\mathbf{e}_y\mathbf{e}_z}+2B_{22}\left(\overbracket{\mathbf{e}_x\mathbf{e}_x}-\overbracket{\mathbf{e}_y\mathbf{e}_y}\right)+4\tilde{B}_{02}\overbracket{\mathbf{e}_x\mathbf{e}_y}\right],\\
\mathbf{S}_\mathcal{D}=&-\frac{8\pi\mu}{a^2}\Big[\frac{5}{3}B_{03}\overbracket{\mathbf{e}_z\mathbf{e}_z\mathbf{e}_z}+B_{13}\left(\overbracket{\mathbf{e}_x\mathbf{e}_y\mathbf{e}_y}-4\overbracket{\mathbf{e}_x\mathbf{e}_z\mathbf{e}_z}+\overbracket{\mathbf{e}_x\mathbf{e}_x\mathbf{e}_x}\right)+\tilde{B}_{13}\left(\overbracket{\mathbf{e}_y\mathbf{e}_x\mathbf{e}_x}-4\overbracket{\mathbf{e}_y\mathbf{e}_z\mathbf{e}_z}+\overbracket{\mathbf{e}_y\mathbf{e}_y\mathbf{e}_y}\right)\nonumber\\
&+10B_{23}\left(\overbracket{\mathbf{e}_x\mathbf{e}_x\mathbf{e}_z}-\overbracket{\mathbf{e}_y\mathbf{e}_y\mathbf{e}_z}\right)+20\tilde{B}_{23}\overbracket{\mathbf{e}_x\mathbf{e}_y\mathbf{e}_z}+10B_{33}\left(3\overbracket{\mathbf{e}_x\mathbf{e}_y\mathbf{e}_y}-\overbracket{\mathbf{e}_x\mathbf{e}_x\mathbf{e}_x}\right)-10\tilde{B}_{33}\left(3\overbracket{\mathbf{e}_y\mathbf{e}_x\mathbf{e}_x}-\overbracket{\mathbf{e}_y\mathbf{e}_y\mathbf{e}_y}\right)\Big],\\
\mathbf{C}_\mathcal{D}=&-\frac{48\pi\mu}{5}\left[C_{02}\overbracket{\mathbf{e}_z\mathbf{e}_z}-2C_{12}\overbracket{\mathbf{e}_x\mathbf{e}_z}-2\tilde{C}_{12}\overbracket{\mathbf{e}_y\mathbf{e}_z}+2C_{22}\left(\overbracket{\mathbf{e}_x\mathbf{e}_x}-\overbracket{\mathbf{e}_y\mathbf{e}_y}\right)+4\tilde{C}_{22}\overbracket{\mathbf{e}_x\mathbf{e}_y}\right],\\
\label{potdipole-pak}
\mathbf{d}=&-\frac{80\pi\mu}{3}\left(B_{01}\mathbf{e}_z -B_{11}\mathbf{e}_x-\tilde{B}_{11}\mathbf{e}_y\right).
\end{align}
By only keeping $B_{0n}$ and $C_{0n}$ terms in equations \eqref{pak1} and \eqref{pak2} and setting the other coefficients to zero, the solution reduces to the axisymmetric motion of a squirmer \citep{pak2014}. In this case, by symmetry, all the force moments generated by the surface activity are invariant by rotation with respect to $\mathbf{e}_{z}$. Thus, as one can see from \eqref{stresslet-pak} to \eqref{potdipole-pak}, they must be of form $\mathbf{e}_z,~\overbracket{\mathbf{e}_z\mathbf{e}_z},~\overbracket{\mathbf{e}_z\mathbf{e}_z\mathbf{e}_z},~\cdots$ which are the irreducible traceless rotation-invariant tensors of $\mathbf{e}_z$. The contribution of non-zero force moments of an axisymmetric squirmer to the flow field is illustrated in Fig.~\ref{flow-field}. Note that, to express the axisymmetric solutions in terms of tangential squirming modes $B_n$ used in \citet{lighthill1952} and \citet{blake1971}, one can simply set $C_{0n}=0$ and substitute $B_{0n}$ by $-\frac{a^{n+2}}{n+1}B_{n}$ in equations \eqref{U-pak} to \eqref{potdipole-pak}.

\subsection{Axisymmetric slender rod}
Let us now consider a slender rod, whose orientation is given by a unit vector $\mathbf{p}$, with an axisymmetric swimming gait along its length $\mathbf{u}^s=\alpha(s)\mathbf{p}$ in an otherwise quiescent fluid. Here $s$ parameterizes (by arclength) the centerline of the rod, e.g. $s\in[-l/2,l/2]$ where $l$ is the length. 

To find the force moments, we first decompose a surface integral into an integration around the perimeter (denoted by $R$) in a plane with normal $\mathbf{p}$ and one over the length of the rod (denoted by $s$) so that $\left<\cdots\right>=\left<\left<\cdots\right>_R\right>_s$. Using the resistive force theory for slender rods \citep{lauga2009a}, we may approximate the force density per unit length
\begin{align}
\left<\mathbf{n}\cdot\hat{\boldsymbol{\sigma}}^\prime\right>_R=-\left[\hat{\zeta}_{\parallel}\mathbf{p}\mathbf{p}+\hat{\zeta}_{\perp}\left(\mathbf{I}-\mathbf{p}\mathbf{p}\right)\right]\cdot\hat{\mathbf{u}}^\prime,
\end{align}
where $\hat{\zeta}_{\parallel}$ and $\hat{\zeta}_{\perp}$ are the parallel and perpendicular drag coefficients. Under this approximation, finding $\hat{\boldsymbol{\mathsf{T}}}$ does not require details of the auxiliary flow field as we illustrate in the following. Recalling that $\hat{\mathbf{u}}^{\prime}=\hat{\boldsymbol{\mathsf{G}}}\odot\hat{\boldsymbol{\mathsf{U}}}^\prime$ and $\hat{\boldsymbol{\sigma}}^{\prime}=\hat{\mu}\hat{\boldsymbol{\mathsf{T}}}\odot\hat{\boldsymbol{\mathsf{U}}}^{\prime}$, one can write
\begin{align}
\label{slender}
\hat{\mu}\left<\mathbf{n}\cdot\hat{\boldsymbol{\mathsf{T}}}\right>=-\left<\left[\hat{\zeta}_{\parallel}\mathbf{p}\mathbf{p}+\hat{\zeta}_{\perp}\left(\mathbf{I}-\mathbf{p}\mathbf{p}\right)\right]\cdot\hat{\boldsymbol{\mathsf{G}}}\right>_s.
\end{align}
We note that $\hat{\boldsymbol{\mathsf{G}}}(\mathbf{x}\in\partial\mathcal{B})$ is known from equations \eqref{U} to \eqref{e}. 

To find the resistance tensors, from equation \eqref{slender} we have
\begin{align}
\hat{\mu}\left<\mathbf{n}\cdot\hat{\mathbf{T}}_U\right>=-\left<\hat{\zeta}_{\parallel}\mathbf{p}\mathbf{p}+\hat{\zeta}_{\perp}\left(\mathbf{I}-\mathbf{p}\mathbf{p}\right)\right>_s,
\end{align}
thus 
\begin{align}
\hat{\mathbf{R}}_{FU}=-\hat{\mu}\left<\mathbf{n}\cdot\hat{\mathbf{T}}_U\right>=\left<\hat{\zeta}_{\parallel}\mathbf{p}\mathbf{p}+\hat{\zeta}_{\perp}\left(\mathbf{I}-\mathbf{p}\mathbf{p}\right)\right>_s=\left[\hat{\zeta}_{\parallel}\mathbf{p}\mathbf{p}+\hat{\zeta}_{\perp} \left(\mathbf{I}-\mathbf{p}\mathbf{p}\right)\right]l.
\end{align}
Similarly, we find $\left[\hat{\mathbf{R}}_{F\Omega},\hat{\mathbf{R}}_{LU},\hat{\mathbf{R}}_{SU},\hat{\mathbf{R}}_{S_\mathcal{D}\Omega},\hat{\mathbf{R}}_{C_\mathcal{D}\Omega},\hat{\mathbf{R}}_{d\Omega}\right]=\mathbf{0}$, $\hat{\mathbf{R}}_{L\Omega}=-\frac{l^3}{12}\hat{\zeta}_{\perp}\left(\mathbf{I}-\mathbf{pp}\right)$, $\hat{\mathbf{R}}_{dU}=\frac{l^3}{12}\left[\hat{\zeta}_{\parallel}\mathbf{pp}+2\hat{\zeta}_{\perp}\left(\mathbf{I}-\mathbf{pp}\right)\right]$ and
\begin{align}
\hat{R}_{S\Omega,ijk}=&\frac{l^3}{24}\hat{\zeta}_{\perp}\left(\epsilon_{iks}p_j+\epsilon_{jks}p_i\right)p_s,\\
\hat{R}_{C_\mathcal{D}U,ijk}=&\frac{l^3}{24}\hat{\zeta}_{\perp}\left(p_i\epsilon_{kjm}+p_j\epsilon_{kim}\right)p_m,\\
\hat{R}_{S_\mathcal{D}U,ijkm}=&\frac{l^3}{12}\left[\hat{\zeta}_\parallel \overbracket{p_ip_jp_k}p_m+\hat{\zeta}_\perp\left(\overbracket{\delta_{im}p_jp_k}^{ijk}-
\overbracket{p_ip_jp_k}p_m\right)\right].
\end{align}

Now to find ${\boldsymbol{\mathsf{S}}}$, we substitute \eqref{slender} in \eqref{babak-eq}. Noting that $\left<{\boldsymbol{\mathsf{D}}}\right>=\mathbf{0}$ since $\left<\mathbf{n}\right>_R=\mathbf{0}$, we arrive at
\begin{align}
\label{babak-eq2}
{\boldsymbol{\mathsf{S}}}&=-\frac{\mu}{\hat{\mu}}\hat{{\boldsymbol{\mathsf{R}}}}\odot\boldsymbol{\mathsf{U}}+\mu\left<\mathbf{u}^s\cdot\left(\mathbf{n}\cdot\hat{\boldsymbol{\mathsf{T}}}\right)\right>,\nonumber\\
&=-\frac{\mu}{\hat{\mu}}\hat{{\boldsymbol{\mathsf{R}}}}\odot\boldsymbol{\mathsf{U}}-\left<\alpha\mathbf{p}\cdot\left[\zeta_{\parallel}\mathbf{p}\mathbf{p}+\zeta_{\perp}\left(\mathbf{I}-\mathbf{p}\mathbf{p}\right)\right]\cdot\hat{\boldsymbol{\mathsf{G}}}\right>_s,\nonumber\\
&=-\frac{\mu}{\hat{\mu}}\hat{{\boldsymbol{\mathsf{R}}}}\odot\boldsymbol{\mathsf{U}}-\zeta_{\parallel}\mathbf{p}\cdot\left<\alpha\hat{\boldsymbol{\mathsf{G}}}\right>_s.
\end{align}
Note that $\zeta_{\parallel}=\left(\mu/\hat{\mu}\right)\hat{\zeta}_{\parallel}$ and $\zeta_{\perp}=\left(\mu/\hat{\mu}\right)\hat{\zeta}_{\perp}$. From this equation, we find $\mathbf{L}^s=\mathbf{0}$, $\mathbf{C}^{s}_{\mathcal{D}}=\mathbf{0}$ and
\begin{align}
\mathbf{F}^s&=-\zeta_{\parallel}\mathbf{p}\cdot\left<\alpha\hat{\mathbf{G}}_U\right>_s=-\zeta_{\parallel}  \left<\alpha\right>_s\mathbf{p},\\
\mathbf{S}^s&=-\zeta_{\parallel}\mathbf{p}\cdot\left<\alpha\hat{\mathbf{G}}_E\right>_s=-\zeta_{\parallel}\left<s\alpha \right>_s\overbracket{\mathbf{pp}},\\
\mathbf{S}^{s}_{\mathcal{D}}&=-\zeta_{\parallel}\mathbf{p}\cdot\left<\alpha\hat{\mathbf{G}}_{\Gamma}\right>_s=-\zeta_{\parallel}\left<s^2\alpha\right>_s\overbracket{\mathbf{ppp}},\\
\mathbf{d}^s&=-\zeta_{\parallel}\mathbf{p}\cdot\left<\alpha\hat{\mathbf{G}}_e\right>_s=-\zeta_{\parallel}\left<s^2\alpha\right>_s\mathbf{p},
\end{align}
which are in the form of irreducible traceless rotation-invariant tensors with regard to symmetry axis $\mathbf{p}$, as expected.

With no external force or torque acting on the rod, we can then determine the translational velocity 
\begin{align}
\mathbf{U}=-(1/l){\left<\mathbf{u}^s\right>_s}=-(1/l){\left<\alpha\right>_s}\mathbf{p},
\end{align}
as also shown by \citet{leshansky2007}. Recalling that $\mathbf{L}^s=\mathbf{0}$, we find $\boldsymbol{\Omega}=\mathbf{0}$.
 
 \section{Conclusion}
In this paper, we investigated the effects of higher-order force moments on the flow field induced by an active particle. Using the boundary integral equations, we expressed the flow as a multipole expansion and decomposed the contribution of second-order force moments into a stresslet dipole, rotlet dipole and a potential dipole. Then, via the reciprocal theorem, we derived explicit formulas for these force moments which are valid for an active particle of arbitrary shape and then evaluated them for a spherical particle, a squirmer and an axisymmetric slender rod. We believe that by providing simple and explicit formulas for more accurate approximations of the flow-fields generated by active particles, we may enhance our understanding of how these particles interact with their surroundings. Given the generality of the employed framework, our results can be extended to capture the effect of third (or higher) order force moments and also can be adapted to study the hydrodynamic interactions between two \citep{sharifi2016} or many \citep{papavassiliou2015} active particles or active particles near boundaries \cite{swan07,spagnolie12}.

\section*{Acknowledgement}
The authors acknowledge funding from the NSERC Grant No. RGPIN-2014-06577.
\appendix*
\section{}
Here we present expressions for the auxiliary flow field and stress field considered for an active spherical particle. For simplicity, we define the flow associated with rigid-body translation as $\hat{\mathbf{u}}^\prime_U=\hat{\mathbf{G}}_U\cdot\left(\hat{\mathbf{U}}-\hat{\mathbf{U}}^\infty\right)$, which from equation \eqref{U} gives $\hat{\mathbf{u}}^\prime_U(\mathbf{x}\in\partial\mathcal{B})=\hat{\mathbf{U}}-\hat{\mathbf{U}}^\infty$. Thus, for a sphere, $\hat{\mathbf{P}}_U$, $\hat{\mathbf{G}}_U$ and $\hat{\mathbf{T}}_U$  can be simply found
\begin{align}
\hat{{P}}_{U,i}&=\frac{3 a}{2\left(x_lx_l\right)^{\frac{3}{2}}}{x}_i,\\
\hat{{G}}_{U,ij}&=\left(\frac{3a}{4\left(x_lx_l\right)^{\frac{1}{2}}}+\frac{a^3}{4\left(x_lx_l\right)^{\frac{3}{2}}}\right)\delta_{ij}+\left(\frac{3a}{4\left(x_lx_l\right)^{\frac{3}{2}}}-\frac{3a^3}{4\left(x_lx_l\right)^{\frac{5}{2}}}\right){x}_i{x}_j,\\
\hat{{T}}_{U,ijk}&=\left(-\frac{9a}{2(x_l x_l)^\frac{5}{2}}+\frac{15 a^3}{2(x_l x_l)^\frac{7}{2}}\right)x_ix_jx_k-\frac{3a^3}{2(x_l x_l)^\frac{5}{2}}\left(x_i \delta_{jk}+x_j \delta_{ik}+x_k \delta_{ij}\right).
\end{align}
Similarly, $\hat{\mathbf{u}}^\prime_\Omega(\mathbf{x}\in\partial\mathcal{B})=\left(\hat{\boldsymbol{\Omega}}-\hat{\boldsymbol{\Omega}}^\infty\right)\times\mathbf{r}$ leads to
\begin{align}
\hat{{P}}_{\Omega,i} &= 0,\\
\hat{{G}}_{\Omega,ij}&=\frac{a^3}{\left(x_lx_l\right)^{\frac{3}{2}}}{\epsilon_{ijk}}{x_k},\\
\hat{{T}}_{\Omega,ijk}&=\frac{3a^3}{(x_l x_l)^\frac{5}{2}}\left[\epsilon_{kis}x_j +\epsilon_{kjs} x_i \right]x_s.
\end{align}
Relevant to the stresslet calculations, we impose $\hat{\mathbf{u}}^\prime_E(\mathbf{x}\in\partial\mathcal{B})=\mathbf{r}\cdot\left(-\hat{\mathbf{E}}^\infty\right)$ in which $\hat{\mathbf{E}}^\infty$ is a symmetric and deviatoric second-order tensor and so
\begin{align}
\hat{{P}}_{E,ij}&= \frac{5 a^3}{\left(x_lx_l\right)^{\frac{5}{2}}} \overbracket{{x}_i{x}_j},\\
\hat{{G}}_{E,ijk}&=\frac{a^5}{\left(x_lx_l\right)^{\frac{5}{2}}} \overbracket{\delta_{ij}x_k}^{jk}+\frac{5}{2} \left(\frac{a^3}{\left(x_lx_l\right)^{\frac{5}{2}}}-\frac{a^5}{\left(x_lx_l\right)^{\frac{7}{2}}}\right) {x}_i\overbracket{{x}_j{x}_k},\\
\hat{T}_{E,ijkm}&=\frac{a^5}{(x_lx_l)^{\frac{5}{2}}}\left(\delta_{kj}\delta_{im} + \delta_{ki}\delta_{jm}\right)+5\left(\frac{7a^5}{(x_lx_l)^{\frac{9}{2}}}-\frac{5a^3}{(x_lx_l)^{\frac{7}{2}}}\right)x_ix_jx_kx_m\nonumber\\
&+\frac{5a^3}{2(x_lx_l)^{\frac{5}{2}}}\left(\delta_{kj} x_i x_m+\delta_{ki} x_j x_m+\delta_{mj} x_i x_k+\delta_{mi} x_j x_k\right)\nonumber\\
&-\frac{5a^5}{(x_lx_l)^{\frac{7}{2}}}\left(\delta_{im} x_k x_j+\delta_{jm} x_k x_i+\delta_{ij} x_k x_m+\delta_{jk} x_i x_m+\delta_{ik} x_jx_m \right).
\end{align}
In determining the stresslet dipole, we have $\hat{\mathbf{u}}_\Gamma^\prime\left(\mathbf{x}\in\partial\mathcal{B}\right)=\mathbf{r}\mathbf{r}:\left(-\hat{\boldsymbol{\Gamma}}^\infty\right)$. Noting that $\hat{\boldsymbol{\Gamma}}^\infty$ is a fully symmetric, deviatoric third-order tensor, we find
\begin{align}
\hat{{P}}_{\Gamma,ijk}&=\frac{35}{4} \frac{a^5}{(x_lx_l)^{\frac{7}{2}}} \overbracket{{x}_i{x}_j{x}_k},\\
\hat{G}_{\Gamma,ijkm}&=\frac{1}{8}\left(\frac{15a^7}{(x_lx_l)^{\frac{7}{2}}}-\frac{7a^5}{(x_lx_l)^{\frac{5}{2}}}\right) \overbracket{\delta_{im}{x}_j{x}_k}^{jkm}+\frac{35}{8} \left(\frac{a^5}{(x_lx_l)^{\frac{7}{2}}}-\frac{a^7}{(x_lx_l)^{\frac{9}{2}}}\right) {x}_i\overbracket{{x}_j{x}_k{x}_m},\\
\hat{{T}}_{\Gamma,ijkmn}&=\frac{15}{8}a^7\left\{\delta_{ni}\left[\frac{\delta_{jk} x_m +  \delta_{mj}x_k}{(x_lx_l)^{\frac{7}{2}}} - \frac{7x_jx_kx_m}{(x_lx_l)^{\frac{9}{2}}}\right] +\delta_{nj}\left[\frac{\delta_{ik} x_m +  \delta_{im}x_k}{(x_lx_l)^{\frac{7}{2}}} - \frac{7x_ix_k x_m}{(x_lx_l)^{\frac{9}{2}}}\right]  \right\}\nonumber\\
&-\frac{7}{8}a^5\left\{\delta_{ni}\left[\frac{\delta_{jk} x_m +  \delta_{mj}x_k}{(x_lx_l)^{\frac{5}{2}}} - \frac{5x_jx_kx_m}{(x_lx_l)^{\frac{7}{2}}}\right] +\delta_{nj}\left[\frac{\delta_{ik} x_m +  \delta_{im}x_k}{(x_lx_l)^{\frac{5}{2}}} - \frac{5x_ix_k x_m}{(x_lx_l)^{\frac{7}{2}}}\right]  \right\}\nonumber\\
&+\frac{35}{4}a^5\left[-\frac{7x_i x_j x_k x_m x_n}{(x_lx_l)^{\frac{9}{2}}}\right]-\frac{35}{4}a^7\left[\frac{\delta_{ij} x_m x_n x_k}{(x_lx_l)^{\frac{9}{2}}}-\frac{9x_i x_j x_m x_n x_k}{(x_lx_l)^{\frac{11}{2}}}\right]\nonumber\\
&+\frac{35}{8}\left(\frac{a^5}{(x_lx_l)^{\frac{7}{2}}} - \frac{a^7}{(x_lx_l)^{\frac{9}{2}}}\right)\left[\delta_{jk}x_i x_m x_n+\delta_{jm}x_i x_k x_n+\delta_{jn}x_i x_k x_m+\delta_{ik}x_j x_m x_n + \delta_{im}x_j x_k x_n + \delta_{in}x_j x_k x_m  \right].
\end{align}
For a sphere with boundary condition $\hat{\mathbf{u}}^\prime_\Lambda(\mathbf{x}\in\partial\mathcal{B})=\left(\boldsymbol{\epsilon}\cdot\mathbf{r}\right)\mathbf{r}:\left(-\hat{\boldsymbol{\Lambda}}^\infty\right)$, wherein $\hat{\boldsymbol{\Lambda}}^\infty$ is a second-order symmetric and deviatoric tensor, we have
\begin{align}
\hat{P}_{\Lambda,ij}&=0,\\
\hat{G}_{\Lambda,ijk}&=a^5\frac{\overbracket{\epsilon_{ijm}x_mx_k}^{jk}}{(x_lx_l)^{\frac{5}{2}}},\\
\hat{T}_{\Lambda,ijkm}&=\frac{a^5}{(x_lx_l)^{\frac{5}{2}}}\left(\epsilon_{iks}\delta_{mj}x_{s}+\epsilon_{jks}\delta_{mi}x_{s}+\epsilon_{kmi}x_{j}+\epsilon_{kmj}x_{i}\right)\nonumber\\
&-\frac{5a^5}{(x_lx_l)^{\frac{7}{2}}}x_mx_s\left(  \epsilon_{iks}x_{j}+   \epsilon_{jms}x_{i}   \right).
\end{align}
Finally, $\hat{\mathbf{u}}_e^\prime(\mathbf{x}\in\partial\mathcal{B})= \left(2|\mathbf{r}|^2\mathbf{I}-\mathbf{r}\mathbf{r}\right)\cdot\left(-\hat{\mathbf{e}}^\infty\right)$ yields
\begin{align}
\hat{P}_{e,i}&= \frac{ 5 a^3}{2(x_lx_l)^{\frac{3}{2}}} {x}_i,\nonumber\\ 
\hat{G}_{e,ij}&=\left(\frac{3a^5}{4(x_lx_l)^{\frac{3}{2}}}+\frac{5a^3}{4(x_lx_l)^{\frac{1}{2}}}\right) \delta_{ij}-\left(\frac{9a^5}{4(x_lx_l)^{\frac{5}{2}}}-\frac{5a^3}{4(x_lx_l)^{\frac{3}{2}}}\right) {x}_i{x}_j,\\
\hat{T}_{e,ijk}&=\frac{15}{2}\left(\frac{3a^5}{(x_lx_l)^\frac{7}{2}}-\frac{a^3}{(x_lx_l)^\frac{5}{2}}\right){x_ix_jx_k}-\frac{9a^5}{2(x_lx_l)^\frac{5}{2}}\left(\delta_{ij}x_k+\delta_{ik}x_j+\delta_{jk}x_i\right).
\end{align}

\bibliography{reference}

\begin{thebibliography}{59}%
\makeatletter
\providecommand \@ifxundefined [1]{%
 \@ifx{#1\undefined}
}%
\providecommand \@ifnum [1]{%
 \ifnum #1\expandafter \@firstoftwo
 \else \expandafter \@secondoftwo
 \fi
}%
\providecommand \@ifx [1]{%
 \ifx #1\expandafter \@firstoftwo
 \else \expandafter \@secondoftwo
 \fi
}%
\providecommand \natexlab [1]{#1}%
\providecommand \enquote  [1]{``#1''}%
\providecommand \bibnamefont  [1]{#1}%
\providecommand \bibfnamefont [1]{#1}%
\providecommand \citenamefont [1]{#1}%
\providecommand \href@noop [0]{\@secondoftwo}%
\providecommand \href [0]{\begingroup \@sanitize@url \@href}%
\providecommand \@href[1]{\@@startlink{#1}\@@href}%
\providecommand \@@href[1]{\endgroup#1\@@endlink}%
\providecommand \@sanitize@url [0]{\catcode `\\12\catcode `\$12\catcode
  `\&12\catcode `\#12\catcode `\^12\catcode `\_12\catcode `\%12\relax}%
\providecommand \@@startlink[1]{}%
\providecommand \@@endlink[0]{}%
\providecommand \url  [0]{\begingroup\@sanitize@url \@url }%
\providecommand \@url [1]{\endgroup\@href {#1}{\urlprefix }}%
\providecommand \urlprefix  [0]{URL }%
\providecommand \Eprint [0]{\href }%
\providecommand \doibase [0]{http://dx.doi.org/}%
\providecommand \selectlanguage [0]{\@gobble}%
\providecommand \bibinfo  [0]{\@secondoftwo}%
\providecommand \bibfield  [0]{\@secondoftwo}%
\providecommand \translation [1]{[#1]}%
\providecommand \BibitemOpen [0]{}%
\providecommand \bibitemStop [0]{}%
\providecommand \bibitemNoStop [0]{.\EOS\space}%
\providecommand \EOS [0]{\spacefactor3000\relax}%
\providecommand \BibitemShut  [1]{\csname bibitem#1\endcsname}%
\let\auto@bib@innerbib\@empty
\bibitem [{\citenamefont {Ramaswamy}(2010)}]{ramaswamy2010}%
  \BibitemOpen
  \bibfield  {author} {\bibinfo {author} {\bibfnamefont {S.}~\bibnamefont
  {Ramaswamy}},\ }\bibfield  {title} {\enquote {\bibinfo {title} {The mechanics
  and statistics of active matter},}\ }\href {\doibase
  10.1146/annurev-conmatphys-070909-104101} {\bibfield  {journal} {\bibinfo
  {journal} {Annu. Rev. Condens. Matter Phys.}\ }\textbf {\bibinfo {volume}
  {1}},\ \bibinfo {pages} {323--345} (\bibinfo {year} {2010})}\BibitemShut
  {NoStop}%
\bibitem [{\citenamefont {Happel}\ and\ \citenamefont
  {Brenner}(1981)}]{happel1981}%
  \BibitemOpen
  \bibfield  {author} {\bibinfo {author} {\bibfnamefont {J.}~\bibnamefont
  {Happel}}\ and\ \bibinfo {author} {\bibfnamefont {H.}~\bibnamefont
  {Brenner}},\ }\href {\doibase 10.1007/978-94-009-8352-6} {\emph {\bibinfo
  {title} {Low {R}eynolds {N}umber {H}ydrodynamics}}}\ (\bibinfo  {publisher}
  {Springer Netherlands},\ \bibinfo {year} {1981})\BibitemShut {NoStop}%
\bibitem [{\citenamefont {Purcell}(1977)}]{purcell1977}%
  \BibitemOpen
  \bibfield  {author} {\bibinfo {author} {\bibfnamefont {E.~M.}\ \bibnamefont
  {Purcell}},\ }\bibfield  {title} {\enquote {\bibinfo {title} {Life at low
  {R}eynolds number},}\ }\href {\doibase 10.1119/1.10903} {\bibfield  {journal}
  {\bibinfo  {journal} {Am. J. Phys.}\ }\textbf {\bibinfo {volume} {45}},\
  \bibinfo {pages} {3--11} (\bibinfo {year} {1977})}\BibitemShut {NoStop}%
\bibitem [{\citenamefont {Lauga}\ and\ \citenamefont
  {Powers}(2009)}]{lauga2009a}%
  \BibitemOpen
  \bibfield  {author} {\bibinfo {author} {\bibfnamefont {E.}~\bibnamefont
  {Lauga}}\ and\ \bibinfo {author} {\bibfnamefont {T.~R.}\ \bibnamefont
  {Powers}},\ }\bibfield  {title} {\enquote {\bibinfo {title} {The
  hydrodynamics of swimming microorganisms},}\ }\href {\doibase
  10.1088/0034-4885/72/9/096601} {\bibfield  {journal} {\bibinfo  {journal}
  {Rep. Prog. Phys.}\ }\textbf {\bibinfo {volume} {72}},\ \bibinfo {pages}
  {096601} (\bibinfo {year} {2009})}\BibitemShut {NoStop}%
\bibitem [{\citenamefont {Lauga}(2011)}]{lauga2011}%
  \BibitemOpen
  \bibfield  {author} {\bibinfo {author} {\bibfnamefont {E.}~\bibnamefont
  {Lauga}},\ }\bibfield  {title} {\enquote {\bibinfo {title} {Life around the
  scallop theorem},}\ }\href {\doibase 10.1039/c0sm00953a} {\bibfield
  {journal} {\bibinfo  {journal} {Soft Matter}\ }\textbf {\bibinfo {volume}
  {7}},\ \bibinfo {pages} {3060--3065} (\bibinfo {year} {2011})}\BibitemShut
  {NoStop}%
\bibitem [{\citenamefont {Nasouri}\ \emph {et~al.}(2017)\citenamefont
  {Nasouri}, \citenamefont {Khot},\ and\ \citenamefont
  {Elfring}}]{nasouri2017}%
  \BibitemOpen
  \bibfield  {author} {\bibinfo {author} {\bibfnamefont {B.}~\bibnamefont
  {Nasouri}}, \bibinfo {author} {\bibfnamefont {A.}~\bibnamefont {Khot}}, \
  and\ \bibinfo {author} {\bibfnamefont {G.~J.}\ \bibnamefont {Elfring}},\
  }\bibfield  {title} {\enquote {\bibinfo {title} {Elastic two-sphere swimmer
  in {S}tokes flow},}\ }\href {\doibase 10.1103/physrevfluids.2.043101}
  {\bibfield  {journal} {\bibinfo  {journal} {Phys. Rev. Fluids}\ }\textbf
  {\bibinfo {volume} {2}},\ \bibinfo {pages} {043101} (\bibinfo {year}
  {2017})}\BibitemShut {NoStop}%
\bibitem [{\citenamefont {Lodish}\ \emph {et~al.}(2000)\citenamefont {Lodish},
  \citenamefont {Berk}, \citenamefont {Zipursky}, \citenamefont {Matsudaira},
  \citenamefont {Baltimore},\ and\ \citenamefont {Darnell}}]{lodish2000}%
  \BibitemOpen
  \bibfield  {author} {\bibinfo {author} {\bibfnamefont {H.}~\bibnamefont
  {Lodish}}, \bibinfo {author} {\bibfnamefont {A.}~\bibnamefont {Berk}},
  \bibinfo {author} {\bibfnamefont {S.~L.}\ \bibnamefont {Zipursky}}, \bibinfo
  {author} {\bibfnamefont {P.}~\bibnamefont {Matsudaira}}, \bibinfo {author}
  {\bibfnamefont {D.}~\bibnamefont {Baltimore}}, \ and\ \bibinfo {author}
  {\bibfnamefont {J.}~\bibnamefont {Darnell}},\ }\href@noop {} {\emph {\bibinfo
  {title} {Cilia and {F}lagella: {S}tructure and {M}ovement}}}\ (\bibinfo
  {publisher} {W. H. Freeman},\ \bibinfo {address} {New York},\ \bibinfo {year}
  {2000})\BibitemShut {NoStop}%
\bibitem [{\citenamefont {Niedermayer}\ \emph {et~al.}(2008)\citenamefont
  {Niedermayer}, \citenamefont {Eckhardt},\ and\ \citenamefont
  {Lenz}}]{niedermayer2008}%
  \BibitemOpen
  \bibfield  {author} {\bibinfo {author} {\bibfnamefont {T.}~\bibnamefont
  {Niedermayer}}, \bibinfo {author} {\bibfnamefont {B.}~\bibnamefont
  {Eckhardt}}, \ and\ \bibinfo {author} {\bibfnamefont {P.}~\bibnamefont
  {Lenz}},\ }\bibfield  {title} {\enquote {\bibinfo {title} {Synchronization,
  phase locking, and metachronal wave formation in ciliary chains},}\ }\href
  {\doibase 10.1063/1.2956984} {\bibfield  {journal} {\bibinfo  {journal}
  {Chaos}\ }\textbf {\bibinfo {volume} {18}},\ \bibinfo {pages} {037128}
  (\bibinfo {year} {2008})}\BibitemShut {NoStop}%
\bibitem [{\citenamefont {Brumley}\ \emph {et~al.}(2012)\citenamefont
  {Brumley}, \citenamefont {Polin}, \citenamefont {Pedley},\ and\ \citenamefont
  {Goldstein}}]{brumley2012}%
  \BibitemOpen
  \bibfield  {author} {\bibinfo {author} {\bibfnamefont {D.~R.}\ \bibnamefont
  {Brumley}}, \bibinfo {author} {\bibfnamefont {M.}~\bibnamefont {Polin}},
  \bibinfo {author} {\bibfnamefont {T.~J.}\ \bibnamefont {Pedley}}, \ and\
  \bibinfo {author} {\bibfnamefont {R.~E.}\ \bibnamefont {Goldstein}},\
  }\bibfield  {title} {\enquote {\bibinfo {title} {Hydrodynamic synchronization
  and metachronal waves on the surface of the colonial alga {V}olvox
  carteri},}\ }\href {\doibase 10.1103/physrevlett.109.268102} {\bibfield
  {journal} {\bibinfo  {journal} {Phys. Rev. Lett.}\ }\textbf {\bibinfo
  {volume} {109}},\ \bibinfo {pages} {268102} (\bibinfo {year}
  {2012})}\BibitemShut {NoStop}%
\bibitem [{\citenamefont {Nasouri}\ and\ \citenamefont
  {Elfring}(2016)}]{nasouri2016}%
  \BibitemOpen
  \bibfield  {author} {\bibinfo {author} {\bibfnamefont {B.}~\bibnamefont
  {Nasouri}}\ and\ \bibinfo {author} {\bibfnamefont {G.~J.}\ \bibnamefont
  {Elfring}},\ }\bibfield  {title} {\enquote {\bibinfo {title} {Hydrodynamic
  interactions of cilia on a spherical body},}\ }\href {\doibase
  10.1103/physreve.93.033111} {\bibfield  {journal} {\bibinfo  {journal} {Phys.
  Rev. E}\ }\textbf {\bibinfo {volume} {93}},\ \bibinfo {pages} {033111}
  (\bibinfo {year} {2016})}\BibitemShut {NoStop}%
\bibitem [{\citenamefont {Quaranta}\ \emph {et~al.}(2015)\citenamefont
  {Quaranta}, \citenamefont {Aubin-Tam},\ and\ \citenamefont
  {Tam}}]{quaranta2015}%
  \BibitemOpen
  \bibfield  {author} {\bibinfo {author} {\bibfnamefont {G.}~\bibnamefont
  {Quaranta}}, \bibinfo {author} {\bibfnamefont {M.~E.}\ \bibnamefont
  {Aubin-Tam}}, \ and\ \bibinfo {author} {\bibfnamefont {D.}~\bibnamefont
  {Tam}},\ }\bibfield  {title} {\enquote {\bibinfo {title} {Hydrodynamics
  versus intracellular coupling in the synchronization of eukaryotic
  flagella},}\ }\href {\doibase 10.1103/physrevlett.115.238101} {\bibfield
  {journal} {\bibinfo  {journal} {Phys. Rev. Lett.}\ }\textbf {\bibinfo
  {volume} {115}},\ \bibinfo {pages} {238101} (\bibinfo {year}
  {2015})}\BibitemShut {NoStop}%
\bibitem [{\citenamefont {Wan}\ and\ \citenamefont
  {Goldstein}(2016)}]{wan2016}%
  \BibitemOpen
  \bibfield  {author} {\bibinfo {author} {\bibfnamefont {K.~Y.}\ \bibnamefont
  {Wan}}\ and\ \bibinfo {author} {\bibfnamefont {R.~E.}\ \bibnamefont
  {Goldstein}},\ }\bibfield  {title} {\enquote {\bibinfo {title} {Coordinated
  beating of algal flagella is mediated by basal coupling},}\ }\href {\doibase
  10.1073/pnas.1518527113} {\bibfield  {journal} {\bibinfo  {journal} {Proc.
  Natl. Acad. Sci.}\ }\textbf {\bibinfo {volume} {113}},\ \bibinfo {pages}
  {E2784--E2793} (\bibinfo {year} {2016})}\BibitemShut {NoStop}%
\bibitem [{\citenamefont {Klindt}\ \emph {et~al.}(2017)\citenamefont {Klindt},
  \citenamefont {Ruloff}, \citenamefont {Wagner},\ and\ \citenamefont
  {Friedrich}}]{klindt2017}%
  \BibitemOpen
  \bibfield  {author} {\bibinfo {author} {\bibfnamefont {G.~S.}\ \bibnamefont
  {Klindt}}, \bibinfo {author} {\bibfnamefont {C.}~\bibnamefont {Ruloff}},
  \bibinfo {author} {\bibfnamefont {C.}~\bibnamefont {Wagner}}, \ and\ \bibinfo
  {author} {\bibfnamefont {B.~M.}\ \bibnamefont {Friedrich}},\ }\bibfield
  {title} {\enquote {\bibinfo {title} {In-phase and anti-phase flagellar
  synchronization by waveform compliance and basal coupling},}\ }\href
  {\doibase 10.1088/1367-2630/aa9031} {\bibfield  {journal} {\bibinfo
  {journal} {New J. Phys.}\ }\textbf {\bibinfo {volume} {19}},\ \bibinfo
  {pages} {113052} (\bibinfo {year} {2017})}\BibitemShut {NoStop}%
\bibitem [{\citenamefont {Anderson}(1989)}]{anderson1989}%
  \BibitemOpen
  \bibfield  {author} {\bibinfo {author} {\bibfnamefont {J.}~\bibnamefont
  {Anderson}},\ }\bibfield  {title} {\enquote {\bibinfo {title} {Colloid
  transport by interfacial forces},}\ }\href {\doibase
  10.1146/annurev.fluid.21.1.61} {\bibfield  {journal} {\bibinfo  {journal}
  {Annu. Rev. Fluid Mech.}\ }\textbf {\bibinfo {volume} {21}},\ \bibinfo
  {pages} {61--99} (\bibinfo {year} {1989})}\BibitemShut {NoStop}%
\bibitem [{\citenamefont {Golestanian}\ \emph {et~al.}(2005)\citenamefont
  {Golestanian}, \citenamefont {Liverpool},\ and\ \citenamefont
  {Ajdari}}]{golestanian2005}%
  \BibitemOpen
  \bibfield  {author} {\bibinfo {author} {\bibfnamefont {R.}~\bibnamefont
  {Golestanian}}, \bibinfo {author} {\bibfnamefont {T.~B.}\ \bibnamefont
  {Liverpool}}, \ and\ \bibinfo {author} {\bibfnamefont {A.}~\bibnamefont
  {Ajdari}},\ }\bibfield  {title} {\enquote {\bibinfo {title} {Propulsion of a
  molecular machine by asymmetric distribution of reaction products},}\ }\href
  {\doibase 10.1103/physrevlett.94.220801} {\bibfield  {journal} {\bibinfo
  {journal} {Phys. Rev. Lett.}\ }\textbf {\bibinfo {volume} {94}},\ \bibinfo
  {pages} {220801} (\bibinfo {year} {2005})}\BibitemShut {NoStop}%
\bibitem [{\citenamefont {Walther}\ and\ \citenamefont
  {M{\"u}ller}(2013)}]{walther2013}%
  \BibitemOpen
  \bibfield  {author} {\bibinfo {author} {\bibfnamefont {A.}~\bibnamefont
  {Walther}}\ and\ \bibinfo {author} {\bibfnamefont {H.~E.}\ \bibnamefont
  {M{\"u}ller}},\ }\bibfield  {title} {\enquote {\bibinfo {title} {Janus
  particles: Synthesis, self-assembly, physical properties, and
  applications},}\ }\href {\doibase 10.1021/cr300089t} {\bibfield  {journal}
  {\bibinfo  {journal} {Chem. Rev.}\ }\textbf {\bibinfo {volume} {113}},\
  \bibinfo {pages} {5194--5261} (\bibinfo {year} {2013})}\BibitemShut {NoStop}%
\bibitem [{\citenamefont {Elgeti}\ \emph {et~al.}(2015)\citenamefont {Elgeti},
  \citenamefont {Winkler},\ and\ \citenamefont {Gompper}}]{elgeti2015}%
  \BibitemOpen
  \bibfield  {author} {\bibinfo {author} {\bibfnamefont {J}~\bibnamefont
  {Elgeti}}, \bibinfo {author} {\bibfnamefont {R.~G.}\ \bibnamefont {Winkler}},
  \ and\ \bibinfo {author} {\bibfnamefont {G.}~\bibnamefont {Gompper}},\
  }\bibfield  {title} {\enquote {\bibinfo {title} {Physics of
  microswimmers{\textemdash}single particle motion and collective behavior: a
  review},}\ }\href {\doibase 10.1088/0034-4885/78/5/056601} {\bibfield
  {journal} {\bibinfo  {journal} {Rep. Prog. Phys.}\ }\textbf {\bibinfo
  {volume} {78}},\ \bibinfo {pages} {056601} (\bibinfo {year}
  {2015})}\BibitemShut {NoStop}%
\bibitem [{\citenamefont {Kim}\ and\ \citenamefont {Karilla}(1991)}]{kim1991}%
  \BibitemOpen
  \bibfield  {author} {\bibinfo {author} {\bibfnamefont {S.}~\bibnamefont
  {Kim}}\ and\ \bibinfo {author} {\bibfnamefont {J.~S.}\ \bibnamefont
  {Karilla}},\ }\href@noop {} {\emph {\bibinfo {title} {Microhydrodynamics:
  principles and selected applications}}}\ (\bibinfo  {publisher}
  {Butterworth-Heinemann},\ \bibinfo {year} {1991})\BibitemShut {NoStop}%
\bibitem [{\citenamefont {Batchelor}(1970)}]{batchelor1970}%
  \BibitemOpen
  \bibfield  {author} {\bibinfo {author} {\bibfnamefont {G.~K.}\ \bibnamefont
  {Batchelor}},\ }\bibfield  {title} {\enquote {\bibinfo {title} {The stress
  system in a suspension of force-free particles},}\ }\href {\doibase
  10.1017/s0022112070000745} {\bibfield  {journal} {\bibinfo  {journal} {J.
  Fluid. Mech.}\ }\textbf {\bibinfo {volume} {41}},\ \bibinfo {pages} {545}
  (\bibinfo {year} {1970})}\BibitemShut {NoStop}%
\bibitem [{\citenamefont {Guell}\ \emph {et~al.}(1988)\citenamefont {Guell},
  \citenamefont {Brenner}, \citenamefont {Frankel},\ and\ \citenamefont
  {Hartman}}]{guell1988}%
  \BibitemOpen
  \bibfield  {author} {\bibinfo {author} {\bibfnamefont {D.~C.}\ \bibnamefont
  {Guell}}, \bibinfo {author} {\bibfnamefont {H.}~\bibnamefont {Brenner}},
  \bibinfo {author} {\bibfnamefont {R.~B.}\ \bibnamefont {Frankel}}, \ and\
  \bibinfo {author} {\bibfnamefont {H.}~\bibnamefont {Hartman}},\ }\bibfield
  {title} {\enquote {\bibinfo {title} {Hydrodynamic forces and band formation
  in swimming magnetotactic bacteria},}\ }\href {\doibase
  10.1016/s0022-5193(88)80274-1} {\bibfield  {journal} {\bibinfo  {journal} {J.
  Theor. Biol.}\ }\textbf {\bibinfo {volume} {135}},\ \bibinfo {pages}
  {525--542} (\bibinfo {year} {1988})}\BibitemShut {NoStop}%
\bibitem [{\citenamefont {Berke}\ \emph {et~al.}(2008)\citenamefont {Berke},
  \citenamefont {Turner}, \citenamefont {Berg},\ and\ \citenamefont
  {Lauga}}]{berke2008}%
  \BibitemOpen
  \bibfield  {author} {\bibinfo {author} {\bibfnamefont {A.~P.}\ \bibnamefont
  {Berke}}, \bibinfo {author} {\bibfnamefont {L.}~\bibnamefont {Turner}},
  \bibinfo {author} {\bibfnamefont {H.~C.}\ \bibnamefont {Berg}}, \ and\
  \bibinfo {author} {\bibfnamefont {E.}~\bibnamefont {Lauga}},\ }\bibfield
  {title} {\enquote {\bibinfo {title} {Hydrodynamic attraction of swimming
  microorganisms by surfaces},}\ }\href {\doibase
  10.1103/physrevlett.101.038102} {\bibfield  {journal} {\bibinfo  {journal}
  {Phys. Rev. Lett.}\ }\textbf {\bibinfo {volume} {101}},\ \bibinfo {pages}
  {038102} (\bibinfo {year} {2008})}\BibitemShut {NoStop}%
\bibitem [{\citenamefont {Lauga}\ and\ \citenamefont
  {Michelin}(2016)}]{lauga2016}%
  \BibitemOpen
  \bibfield  {author} {\bibinfo {author} {\bibfnamefont {E.}~\bibnamefont
  {Lauga}}\ and\ \bibinfo {author} {\bibfnamefont {S.}~\bibnamefont
  {Michelin}},\ }\bibfield  {title} {\enquote {\bibinfo {title} {Stresslets
  induced by active swimmers},}\ }\href {\doibase
  10.1103/physrevlett.117.148001} {\bibfield  {journal} {\bibinfo  {journal}
  {Phys. Rev. Lett.}\ }\textbf {\bibinfo {volume} {117}},\ \bibinfo {pages}
  {148001} (\bibinfo {year} {2016})}\BibitemShut {NoStop}%
\bibitem [{\citenamefont {Saintillan}(2009)}]{saintillan2009}%
  \BibitemOpen
  \bibfield  {author} {\bibinfo {author} {\bibfnamefont {D.}~\bibnamefont
  {Saintillan}},\ }\bibfield  {title} {\enquote {\bibinfo {title} {The dilute
  rheology of swimming suspensions: A simple kinetic model},}\ }\href {\doibase
  10.1007/s11340-009-9267-0} {\bibfield  {journal} {\bibinfo  {journal} {Exp.
  Mech}\ }\textbf {\bibinfo {volume} {50}},\ \bibinfo {pages} {1275--1281}
  (\bibinfo {year} {2009})}\BibitemShut {NoStop}%
\bibitem [{\citenamefont {Saintillan}\ and\ \citenamefont
  {Shelley}(2013)}]{saintillan2013}%
  \BibitemOpen
  \bibfield  {author} {\bibinfo {author} {\bibfnamefont {D.}~\bibnamefont
  {Saintillan}}\ and\ \bibinfo {author} {\bibfnamefont {M.~J.}\ \bibnamefont
  {Shelley}},\ }\bibfield  {title} {\enquote {\bibinfo {title} {Active
  suspensions and their nonlinear models},}\ }\href {\doibase
  10.1016/j.crhy.2013.04.001} {\bibfield  {journal} {\bibinfo  {journal} {C. R.
  Physique}\ }\textbf {\bibinfo {volume} {14}},\ \bibinfo {pages} {497--517}
  (\bibinfo {year} {2013})}\BibitemShut {NoStop}%
\bibitem [{\citenamefont {Dombrowski}\ \emph {et~al.}(2004)\citenamefont
  {Dombrowski}, \citenamefont {Cisneros}, \citenamefont {Chatkaew},
  \citenamefont {Goldstein},\ and\ \citenamefont {Kessler}}]{dombrowski2004}%
  \BibitemOpen
  \bibfield  {author} {\bibinfo {author} {\bibfnamefont {C.}~\bibnamefont
  {Dombrowski}}, \bibinfo {author} {\bibfnamefont {L.}~\bibnamefont
  {Cisneros}}, \bibinfo {author} {\bibfnamefont {S.}~\bibnamefont {Chatkaew}},
  \bibinfo {author} {\bibfnamefont {R.~E.}\ \bibnamefont {Goldstein}}, \ and\
  \bibinfo {author} {\bibfnamefont {J.~O.}\ \bibnamefont {Kessler}},\
  }\bibfield  {title} {\enquote {\bibinfo {title} {Self-concentration and
  large-scale coherence in bacterial dynamics},}\ }\href {\doibase
  10.1103/physrevlett.93.098103} {\bibfield  {journal} {\bibinfo  {journal}
  {Phys. Rev. Lett.}\ }\textbf {\bibinfo {volume} {93}},\ \bibinfo {pages}
  {098103} (\bibinfo {year} {2004})}\BibitemShut {NoStop}%
\bibitem [{\citenamefont {Drescher}\ \emph {et~al.}(2009)\citenamefont
  {Drescher}, \citenamefont {Leptos}, \citenamefont {Tuval}, \citenamefont
  {Ishikawa}, \citenamefont {Pedley},\ and\ \citenamefont
  {Goldstein}}]{drescher2009}%
  \BibitemOpen
  \bibfield  {author} {\bibinfo {author} {\bibfnamefont {K.}~\bibnamefont
  {Drescher}}, \bibinfo {author} {\bibfnamefont {K.~C.}\ \bibnamefont
  {Leptos}}, \bibinfo {author} {\bibfnamefont {I.}~\bibnamefont {Tuval}},
  \bibinfo {author} {\bibfnamefont {T.}~\bibnamefont {Ishikawa}}, \bibinfo
  {author} {\bibfnamefont {T.~J.}\ \bibnamefont {Pedley}}, \ and\ \bibinfo
  {author} {\bibfnamefont {R.~E.}\ \bibnamefont {Goldstein}},\ }\bibfield
  {title} {\enquote {\bibinfo {title} {{Dancing \textit{Volvox}}: Hydrodynamic
  bound states of swimming algae},}\ }\href {\doibase
  10.1103/physrevlett.102.168101} {\bibfield  {journal} {\bibinfo  {journal}
  {Phys. Rev. Lett.}\ }\textbf {\bibinfo {volume} {102}},\ \bibinfo {pages}
  {168101} (\bibinfo {year} {2009})}\BibitemShut {NoStop}%
\bibitem [{\citenamefont {Drescher}\ \emph {et~al.}(2010)\citenamefont
  {Drescher}, \citenamefont {Goldstein}, \citenamefont {Michel}, \citenamefont
  {Polin},\ and\ \citenamefont {Tuval}}]{drescher2010}%
  \BibitemOpen
  \bibfield  {author} {\bibinfo {author} {\bibfnamefont {K.}~\bibnamefont
  {Drescher}}, \bibinfo {author} {\bibfnamefont {R.~E.}\ \bibnamefont
  {Goldstein}}, \bibinfo {author} {\bibfnamefont {N.}~\bibnamefont {Michel}},
  \bibinfo {author} {\bibfnamefont {M.}~\bibnamefont {Polin}}, \ and\ \bibinfo
  {author} {\bibfnamefont {I.}~\bibnamefont {Tuval}},\ }\bibfield  {title}
  {\enquote {\bibinfo {title} {Direct measurement of the flow field around
  swimming microorganisms},}\ }\href {\doibase 10.1103/PhysRevLett.105.168101}
  {\bibfield  {journal} {\bibinfo  {journal} {Phys. Rev. Lett.}\ }\textbf
  {\bibinfo {volume} {105}},\ \bibinfo {pages} {168101} (\bibinfo {year}
  {2010})}\BibitemShut {NoStop}%
\bibitem [{\citenamefont {Guasto}\ \emph {et~al.}(2010)\citenamefont {Guasto},
  \citenamefont {Johnson},\ and\ \citenamefont {Gollub}}]{guasto2010}%
  \BibitemOpen
  \bibfield  {author} {\bibinfo {author} {\bibfnamefont {J.~S.}\ \bibnamefont
  {Guasto}}, \bibinfo {author} {\bibfnamefont {K.~A.}\ \bibnamefont {Johnson}},
  \ and\ \bibinfo {author} {\bibfnamefont {J.~P.}\ \bibnamefont {Gollub}},\
  }\bibfield  {title} {\enquote {\bibinfo {title} {Direct measurement of the
  flow field around swimming microorganisms},}\ }\href {\doibase
  10.1103/PhysRevLett.105.168102} {\bibfield  {journal} {\bibinfo  {journal}
  {Phys. Rev. Lett.}\ }\textbf {\bibinfo {volume} {105}},\ \bibinfo {pages}
  {168102} (\bibinfo {year} {2010})}\BibitemShut {NoStop}%
\bibitem [{\citenamefont {Ghose}\ and\ \citenamefont
  {Adhikari}(2014)}]{ghose2014}%
  \BibitemOpen
  \bibfield  {author} {\bibinfo {author} {\bibfnamefont {S.}~\bibnamefont
  {Ghose}}\ and\ \bibinfo {author} {\bibfnamefont {R.}~\bibnamefont
  {Adhikari}},\ }\bibfield  {title} {\enquote {\bibinfo {title} {Irreducible
  representations of oscillatory and swirling flows in active soft matter},}\
  }\href {\doibase 10.1103/physrevlett.112.118102} {\bibfield  {journal}
  {\bibinfo  {journal} {Phys. Rev. Lett.}\ }\textbf {\bibinfo {volume} {112}},\
  \bibinfo {pages} {118102} (\bibinfo {year} {2014})}\BibitemShut {NoStop}%
\bibitem [{\citenamefont {Elfring}(2017)}]{elfring2017}%
  \BibitemOpen
  \bibfield  {author} {\bibinfo {author} {\bibfnamefont {G.~J.}\ \bibnamefont
  {Elfring}},\ }\bibfield  {title} {\enquote {\bibinfo {title} {Force moments
  of an active particle in a complex fluid},}\ }\href {\doibase
  10.1017/jfm.2017.632} {\bibfield  {journal} {\bibinfo  {journal} {J. Fluid.
  Mech.}\ }\textbf {\bibinfo {volume} {829}},\ \bibinfo {pages} {R3} (\bibinfo
  {year} {2017})}\BibitemShut {NoStop}%
\bibitem [{\citenamefont {Hinch}(1972)}]{hinch1972}%
  \BibitemOpen
  \bibfield  {author} {\bibinfo {author} {\bibfnamefont {E.~J.}\ \bibnamefont
  {Hinch}},\ }\bibfield  {title} {\enquote {\bibinfo {title} {Note on the
  symmetries of certain material tensors for a particle in {S}tokes flow},}\
  }\href {\doibase 10.1017/s0022112072000771} {\bibfield  {journal} {\bibinfo
  {journal} {J. Fluid. Mech.}\ }\textbf {\bibinfo {volume} {54}},\ \bibinfo
  {pages} {423} (\bibinfo {year} {1972})}\BibitemShut {NoStop}%
\bibitem [{\citenamefont {Rallison}(1978)}]{rallison1978}%
  \BibitemOpen
  \bibfield  {author} {\bibinfo {author} {\bibfnamefont {J.~M.}\ \bibnamefont
  {Rallison}},\ }\bibfield  {title} {\enquote {\bibinfo {title} {Note on the
  {F}ax\'en relations for a particle in {S}tokes flow},}\ }\href {\doibase
  10.1017/s0022112078002256} {\bibfield  {journal} {\bibinfo  {journal} {J.
  Fluid. Mech.}\ }\textbf {\bibinfo {volume} {88}},\ \bibinfo {pages} {529}
  (\bibinfo {year} {1978})}\BibitemShut {NoStop}%
\bibitem [{\citenamefont {Leal}(1979)}]{leal1979}%
  \BibitemOpen
  \bibfield  {author} {\bibinfo {author} {\bibfnamefont {L.~G.}\ \bibnamefont
  {Leal}},\ }\bibfield  {title} {\enquote {\bibinfo {title} {The motion of
  small particles in non-{N}ewtonian fluids},}\ }\href {\doibase
  10.1016/0377-0257(79)85004-1} {\bibfield  {journal} {\bibinfo  {journal} {J.
  Non-Newton. Fluid Mech.}\ }\textbf {\bibinfo {volume} {5}},\ \bibinfo {pages}
  {33--78} (\bibinfo {year} {1979})}\BibitemShut {NoStop}%
\bibitem [{\citenamefont {Spagnolie}\ and\ \citenamefont
  {Lauga}(2010)}]{spagnolie2010}%
  \BibitemOpen
  \bibfield  {author} {\bibinfo {author} {\bibfnamefont {S.~E.}\ \bibnamefont
  {Spagnolie}}\ and\ \bibinfo {author} {\bibfnamefont {E.}~\bibnamefont
  {Lauga}},\ }\bibfield  {title} {\enquote {\bibinfo {title} {Jet propulsion
  without inertia},}\ }\href {\doibase 10.1063/1.3469786} {\bibfield  {journal}
  {\bibinfo  {journal} {Phys. Fluids}\ }\textbf {\bibinfo {volume} {22}},\
  \bibinfo {pages} {081902} (\bibinfo {year} {2010})}\BibitemShut {NoStop}%
\bibitem [{\citenamefont {Michelin}\ and\ \citenamefont
  {Lauga}(2015)}]{michelin2015}%
  \BibitemOpen
  \bibfield  {author} {\bibinfo {author} {\bibfnamefont {S.}~\bibnamefont
  {Michelin}}\ and\ \bibinfo {author} {\bibfnamefont {E.}~\bibnamefont
  {Lauga}},\ }\bibfield  {title} {\enquote {\bibinfo {title} {A reciprocal
  theorem for boundary-driven channel flows},}\ }\href {\doibase
  10.1063/1.4935415} {\bibfield  {journal} {\bibinfo  {journal} {Phys. Fluids}\
  }\textbf {\bibinfo {volume} {27}},\ \bibinfo {pages} {111701} (\bibinfo
  {year} {2015})}\BibitemShut {NoStop}%
\bibitem [{\citenamefont {Pak}\ \emph {et~al.}(2014)\citenamefont {Pak},
  \citenamefont {Feng},\ and\ \citenamefont {Stone}}]{pak2014b}%
  \BibitemOpen
  \bibfield  {author} {\bibinfo {author} {\bibfnamefont {O.~S.}\ \bibnamefont
  {Pak}}, \bibinfo {author} {\bibfnamefont {J.}~\bibnamefont {Feng}}, \ and\
  \bibinfo {author} {\bibfnamefont {H.~A.}\ \bibnamefont {Stone}},\ }\bibfield
  {title} {\enquote {\bibinfo {title} {Viscous marangoni migration of a drop in
  a poiseuille flow at low surface p{\'{e}}clet numbers},}\ }\href {\doibase
  10.1017/jfm.2014.380} {\bibfield  {journal} {\bibinfo  {journal} {J. Fluid.
  Mech.}\ }\textbf {\bibinfo {volume} {753}},\ \bibinfo {pages} {535--552}
  (\bibinfo {year} {2014})}\BibitemShut {NoStop}%
\bibitem [{\citenamefont {Stone}\ and\ \citenamefont
  {Samuel}(1996)}]{stone1996}%
  \BibitemOpen
  \bibfield  {author} {\bibinfo {author} {\bibfnamefont {H.~A.}\ \bibnamefont
  {Stone}}\ and\ \bibinfo {author} {\bibfnamefont {A.~D.~T.}\ \bibnamefont
  {Samuel}},\ }\bibfield  {title} {\enquote {\bibinfo {title} {Propulsion of
  microorganisms by surface distortions},}\ }\href {\doibase
  10.1103/physrevlett.77.4102} {\bibfield  {journal} {\bibinfo  {journal}
  {Phys. Rev. Lett.}\ }\textbf {\bibinfo {volume} {77}},\ \bibinfo {pages}
  {4102--4104} (\bibinfo {year} {1996})}\BibitemShut {NoStop}%
\bibitem [{\citenamefont {Elfring}(2015)}]{elfring2015}%
  \BibitemOpen
  \bibfield  {author} {\bibinfo {author} {\bibfnamefont {G.~J.}\ \bibnamefont
  {Elfring}},\ }\bibfield  {title} {\enquote {\bibinfo {title} {A note on the
  reciprocal theorem for the swimming of simple bodies},}\ }\href {\doibase
  10.1063/1.4906993} {\bibfield  {journal} {\bibinfo  {journal} {Phys. Fluids}\
  }\textbf {\bibinfo {volume} {27}},\ \bibinfo {pages} {023101} (\bibinfo
  {year} {2015})}\BibitemShut {NoStop}%
\bibitem [{\citenamefont {Lauga}(2009)}]{lauga09}%
  \BibitemOpen
  \bibfield  {author} {\bibinfo {author} {\bibfnamefont {E.}~\bibnamefont
  {Lauga}},\ }\bibfield  {title} {\enquote {\bibinfo {title} {Life at high
  {D}eborah number},}\ }\href@noop {} {\bibfield  {journal} {\bibinfo
  {journal} {Europhys. Lett.}\ }\textbf {\bibinfo {volume} {86}},\ \bibinfo
  {pages} {64001} (\bibinfo {year} {2009})}\BibitemShut {NoStop}%
\bibitem [{\citenamefont {Pak}\ \emph {et~al.}(2012)\citenamefont {Pak},
  \citenamefont {Zhu}, \citenamefont {Brandt},\ and\ \citenamefont
  {Lauga}}]{pak12}%
  \BibitemOpen
  \bibfield  {author} {\bibinfo {author} {\bibfnamefont {O.~S.}\ \bibnamefont
  {Pak}}, \bibinfo {author} {\bibfnamefont {L.}~\bibnamefont {Zhu}}, \bibinfo
  {author} {\bibfnamefont {L.}~\bibnamefont {Brandt}}, \ and\ \bibinfo {author}
  {\bibfnamefont {E.}~\bibnamefont {Lauga}},\ }\bibfield  {title} {\enquote
  {\bibinfo {title} {Micropropulsion and microrheology in complex fluids via
  symmetry breaking},}\ }\href@noop {} {\bibfield  {journal} {\bibinfo
  {journal} {Phys. Fluids}\ }\textbf {\bibinfo {volume} {24}},\ \bibinfo
  {pages} {103102} (\bibinfo {year} {2012})}\BibitemShut {NoStop}%
\bibitem [{\citenamefont {Lauga}(2014)}]{lauga14}%
  \BibitemOpen
  \bibfield  {author} {\bibinfo {author} {\bibfnamefont {E.}~\bibnamefont
  {Lauga}},\ }\bibfield  {title} {\enquote {\bibinfo {title} {Locomotion in
  complex fluids: Integral theorems},}\ }\href {\doibase 10.1063/1.4891969}
  {\bibfield  {journal} {\bibinfo  {journal} {Phys. Fluids}\ }\textbf {\bibinfo
  {volume} {26}},\ \bibinfo {pages} {081902} (\bibinfo {year}
  {2014})}\BibitemShut {NoStop}%
\bibitem [{\citenamefont {Datt}\ \emph {et~al.}(2015)\citenamefont {Datt},
  \citenamefont {Zhu}, \citenamefont {Elfring},\ and\ \citenamefont
  {Pak}}]{datt2015}%
  \BibitemOpen
  \bibfield  {author} {\bibinfo {author} {\bibfnamefont {C.}~\bibnamefont
  {Datt}}, \bibinfo {author} {\bibfnamefont {L.}~\bibnamefont {Zhu}}, \bibinfo
  {author} {\bibfnamefont {G.~J.}\ \bibnamefont {Elfring}}, \ and\ \bibinfo
  {author} {\bibfnamefont {O.~S.}\ \bibnamefont {Pak}},\ }\bibfield  {title}
  {\enquote {\bibinfo {title} {Squirming through shear-thinning fluids},}\
  }\href {\doibase 10.1017/jfm.2015.600} {\bibfield  {journal} {\bibinfo
  {journal} {J. Fluid. Mech.}\ }\textbf {\bibinfo {volume} {784}},\ \bibinfo
  {pages} {R1} (\bibinfo {year} {2015})}\BibitemShut {NoStop}%
\bibitem [{\citenamefont {Datt}\ \emph {et~al.}(2017)\citenamefont {Datt},
  \citenamefont {Natale}, \citenamefont {Hatzikiriakos},\ and\ \citenamefont
  {Elfring}}]{datt2017}%
  \BibitemOpen
  \bibfield  {author} {\bibinfo {author} {\bibfnamefont {C.}~\bibnamefont
  {Datt}}, \bibinfo {author} {\bibfnamefont {G.}~\bibnamefont {Natale}},
  \bibinfo {author} {\bibfnamefont {S.~G.}\ \bibnamefont {Hatzikiriakos}}, \
  and\ \bibinfo {author} {\bibfnamefont {G.~J.}\ \bibnamefont {Elfring}},\
  }\bibfield  {title} {\enquote {\bibinfo {title} {An active particle in a
  complex fluid},}\ }\href {\doibase 10.1017/jfm.2017.353} {\bibfield
  {journal} {\bibinfo  {journal} {J. Fluid. Mech.}\ }\textbf {\bibinfo {volume}
  {823}},\ \bibinfo {pages} {675--688} (\bibinfo {year} {2017})}\BibitemShut
  {NoStop}%
\bibitem [{\citenamefont {Pozrikidis}(1992)}]{pozrikidis1992}%
  \BibitemOpen
  \bibfield  {author} {\bibinfo {author} {\bibfnamefont {C.}~\bibnamefont
  {Pozrikidis}},\ }\href {\doibase https://doi.org/10.1017/CBO9780511624124}
  {\emph {\bibinfo {title} {Boundary integral and singularity methods for
  linearized viscous flow}}}\ (\bibinfo  {publisher} {Cambridge University
  Press},\ \bibinfo {year} {1992})\BibitemShut {NoStop}%
\bibitem [{\citenamefont {Pozrikidis}(2011)}]{pozrikidis2011}%
  \BibitemOpen
  \bibfield  {author} {\bibinfo {author} {\bibfnamefont {C.}~\bibnamefont
  {Pozrikidis}},\ }\href@noop {} {\emph {\bibinfo {title} {Introduction to
  theoretical and computational fluid dynamics}}}\ (\bibinfo  {publisher}
  {Oxford University Press},\ \bibinfo {year} {2011})\BibitemShut {NoStop}%
\bibitem [{\citenamefont {Andrews}\ and\ \citenamefont
  {Wilkes}(1985)}]{andrews1985}%
  \BibitemOpen
  \bibfield  {author} {\bibinfo {author} {\bibfnamefont {D.~L.}\ \bibnamefont
  {Andrews}}\ and\ \bibinfo {author} {\bibfnamefont {P.~J.}\ \bibnamefont
  {Wilkes}},\ }\bibfield  {title} {\enquote {\bibinfo {title} {Irreducible
  tensors and selection rules for three-frequency absorption},}\ }\href
  {\doibase 10.1063/1.449343} {\bibfield  {journal} {\bibinfo  {journal} {J.
  Chem. Phys.}\ }\textbf {\bibinfo {volume} {83}},\ \bibinfo {pages}
  {2009--2014} (\bibinfo {year} {1985})}\BibitemShut {NoStop}%
\bibitem [{\citenamefont {Andrews}\ \emph {et~al.}(1988)\citenamefont
  {Andrews}, \citenamefont {Blake},\ and\ \citenamefont
  {Hopkins}}]{andrews1988}%
  \BibitemOpen
  \bibfield  {author} {\bibinfo {author} {\bibfnamefont {D.~L.}\ \bibnamefont
  {Andrews}}, \bibinfo {author} {\bibfnamefont {N.~P.}\ \bibnamefont {Blake}},
  \ and\ \bibinfo {author} {\bibfnamefont {K.~P.}\ \bibnamefont {Hopkins}},\
  }\bibfield  {title} {\enquote {\bibinfo {title} {Theory of electro-optical
  effects in two-photon spectroscopy},}\ }\href {\doibase 10.1063/1.454494}
  {\bibfield  {journal} {\bibinfo  {journal} {J. Chem. Phys.}\ }\textbf
  {\bibinfo {volume} {88}},\ \bibinfo {pages} {6022--6029} (\bibinfo {year}
  {1988})}\BibitemShut {NoStop}%
\bibitem [{\citenamefont {Spagnolie}\ and\ \citenamefont
  {Lauga}(2012)}]{spagnolie12}%
  \BibitemOpen
  \bibfield  {author} {\bibinfo {author} {\bibfnamefont {S.~E.}\ \bibnamefont
  {Spagnolie}}\ and\ \bibinfo {author} {\bibfnamefont {E.}~\bibnamefont
  {Lauga}},\ }\bibfield  {title} {\enquote {\bibinfo {title} {Hydrodynamics of
  self-propulsion near a boundary: predictions and accuracy of far-field
  approximations},}\ }\href {\doibase 10.1017/jfm.2012.101} {\bibfield
  {journal} {\bibinfo  {journal} {J. Fluid Mech}\ }\textbf {\bibinfo {volume}
  {700}},\ \bibinfo {pages} {105–147} (\bibinfo {year} {2012})}\BibitemShut
  {NoStop}%
\bibitem [{\citenamefont {Smith}\ and\ \citenamefont
  {Blake}(2009)}]{smith2009}%
  \BibitemOpen
  \bibfield  {author} {\bibinfo {author} {\bibfnamefont {D.~J.}\ \bibnamefont
  {Smith}}\ and\ \bibinfo {author} {\bibfnamefont {J.~R.}\ \bibnamefont
  {Blake}},\ }\bibfield  {title} {\enquote {\bibinfo {title} {Surface
  accumulation of spermatozoa: A fluid dynamic phenomenon.}}\ }\href@noop {}
  {\bibfield  {journal} {\bibinfo  {journal} {Math. Sci.}\ }\textbf {\bibinfo
  {volume} {34}},\ \bibinfo {pages} {74--87} (\bibinfo {year}
  {2009})}\BibitemShut {NoStop}%
\bibitem [{\citenamefont {Michelin}\ and\ \citenamefont
  {Lauga}(2014)}]{michelin14}%
  \BibitemOpen
  \bibfield  {author} {\bibinfo {author} {\bibfnamefont {S.}~\bibnamefont
  {Michelin}}\ and\ \bibinfo {author} {\bibfnamefont {E.}~\bibnamefont
  {Lauga}},\ }\bibfield  {title} {\enquote {\bibinfo {title} {Phoretic
  self-propulsion at finite {P}{\`e}clet numbers},}\ }\href {\doibase
  10.1017/jfm.2014.158} {\bibfield  {journal} {\bibinfo  {journal} {J. Fluid
  Mech.}\ }\textbf {\bibinfo {volume} {747}},\ \bibinfo {pages} {572 -- 604}
  (\bibinfo {year} {2014})}\BibitemShut {NoStop}%
\bibitem [{\citenamefont {Elfring}\ and\ \citenamefont
  {Lauga}(2015)}]{elfring15}%
  \BibitemOpen
  \bibfield  {author} {\bibinfo {author} {\bibfnamefont {G.~J.}\ \bibnamefont
  {Elfring}}\ and\ \bibinfo {author} {\bibfnamefont {E.}~\bibnamefont
  {Lauga}},\ }\bibfield  {title} {\enquote {\bibinfo {title} {Theory of
  locomotion in complex fluids},}\ }in\ \href@noop {} {\emph {\bibinfo
  {booktitle} {Complex Fluids in Biological Systems}}}\ (\bibinfo  {publisher}
  {Springer},\ \bibinfo {year} {2015})\ pp.\ \bibinfo {pages}
  {285--319}\BibitemShut {NoStop}%
\bibitem [{\citenamefont {Anderson}\ and\ \citenamefont
  {Prieve}(1991)}]{anderson91}%
  \BibitemOpen
  \bibfield  {author} {\bibinfo {author} {\bibfnamefont {J.~L.}\ \bibnamefont
  {Anderson}}\ and\ \bibinfo {author} {\bibfnamefont {D.~C.}\ \bibnamefont
  {Prieve}},\ }\bibfield  {title} {\enquote {\bibinfo {title} {Diffusiophoresis
  caused by gradients of strongly adsorbing solutes},}\ }\href {\doibase
  10.1021/la00050a035} {\bibfield  {journal} {\bibinfo  {journal} {Langmuir}\
  }\textbf {\bibinfo {volume} {7}},\ \bibinfo {pages} {403--406} (\bibinfo
  {year} {1991})}\BibitemShut {NoStop}%
\bibitem [{\citenamefont {Lighthill}(1952)}]{lighthill1952}%
  \BibitemOpen
  \bibfield  {author} {\bibinfo {author} {\bibfnamefont {M.~J.}\ \bibnamefont
  {Lighthill}},\ }\bibfield  {title} {\enquote {\bibinfo {title} {On the
  squirming motion of nearly spherical deformable bodies through liquids at
  very small {Reynolds} numbers},}\ }\href {\doibase 10.1002/cpa.3160050201}
  {\bibfield  {journal} {\bibinfo  {journal} {Comm. Pure Appl. Math}\ }\textbf
  {\bibinfo {volume} {5}},\ \bibinfo {pages} {109--118} (\bibinfo {year}
  {1952})}\BibitemShut {NoStop}%
\bibitem [{\citenamefont {Blake}(1971)}]{blake1971}%
  \BibitemOpen
  \bibfield  {author} {\bibinfo {author} {\bibfnamefont {J.~R.}\ \bibnamefont
  {Blake}},\ }\bibfield  {title} {\enquote {\bibinfo {title} {A spherical
  envelope approach to ciliary propulsion},}\ }\href {\doibase
  10.1017/S002211207100048X} {\bibfield  {journal} {\bibinfo  {journal} {J.
  Fluid Mech.}\ }\textbf {\bibinfo {volume} {46}},\ \bibinfo {pages} {199--208}
  (\bibinfo {year} {1971})}\BibitemShut {NoStop}%
\bibitem [{\citenamefont {Pak}\ and\ \citenamefont {Lauga}(2014)}]{pak2014}%
  \BibitemOpen
  \bibfield  {author} {\bibinfo {author} {\bibfnamefont {O.~S.}\ \bibnamefont
  {Pak}}\ and\ \bibinfo {author} {\bibfnamefont {E.}~\bibnamefont {Lauga}},\
  }\bibfield  {title} {\enquote {\bibinfo {title} {Generalized squirming motion
  of a sphere},}\ }\href {\doibase 10.1007/s10665-014-9690-9} {\bibfield
  {journal} {\bibinfo  {journal} {J. Engrg. Math.}\ }\textbf {\bibinfo {volume}
  {88}},\ \bibinfo {pages} {1--28} (\bibinfo {year} {2014})}\BibitemShut
  {NoStop}%
\bibitem [{\citenamefont {Leshansky}\ \emph {et~al.}(2007)\citenamefont
  {Leshansky}, \citenamefont {Kenneth}, \citenamefont {Gat},\ and\
  \citenamefont {Avron}}]{leshansky2007}%
  \BibitemOpen
  \bibfield  {author} {\bibinfo {author} {\bibfnamefont {A.~M.}\ \bibnamefont
  {Leshansky}}, \bibinfo {author} {\bibfnamefont {O.}~\bibnamefont {Kenneth}},
  \bibinfo {author} {\bibfnamefont {O.}~\bibnamefont {Gat}}, \ and\ \bibinfo
  {author} {\bibfnamefont {J.~E.}\ \bibnamefont {Avron}},\ }\bibfield  {title}
  {\enquote {\bibinfo {title} {A frictionless microswimmer},}\ }\href {\doibase
  10.1088/1367-2630/9/5/145} {\bibfield  {journal} {\bibinfo  {journal} {New J.
  Phys.}\ }\textbf {\bibinfo {volume} {9}},\ \bibinfo {pages} {145--145}
  (\bibinfo {year} {2007})}\BibitemShut {NoStop}%
\bibitem [{\citenamefont {Sharifi-Mood}\ \emph {et~al.}(2016)\citenamefont
  {Sharifi-Mood}, \citenamefont {Mozaffari},\ and\ \citenamefont
  {C{\'{o}}rdova-Figueroa}}]{sharifi2016}%
  \BibitemOpen
  \bibfield  {author} {\bibinfo {author} {\bibfnamefont {N.}~\bibnamefont
  {Sharifi-Mood}}, \bibinfo {author} {\bibfnamefont {A.}~\bibnamefont
  {Mozaffari}}, \ and\ \bibinfo {author} {\bibfnamefont {U.~M.}\ \bibnamefont
  {C{\'{o}}rdova-Figueroa}},\ }\bibfield  {title} {\enquote {\bibinfo {title}
  {Pair interaction of catalytically active colloids: from assembly to
  escape},}\ }\href {\doibase 10.1017/jfm.2016.317} {\bibfield  {journal}
  {\bibinfo  {journal} {J. Fluid. Mech.}\ }\textbf {\bibinfo {volume} {798}},\
  \bibinfo {pages} {910--954} (\bibinfo {year} {2016})}\BibitemShut {NoStop}%
\bibitem [{\citenamefont {Papavassiliou}\ and\ \citenamefont
  {Alexander}(2015)}]{papavassiliou2015}%
  \BibitemOpen
  \bibfield  {author} {\bibinfo {author} {\bibfnamefont {D.}~\bibnamefont
  {Papavassiliou}}\ and\ \bibinfo {author} {\bibfnamefont {G.~P.}\ \bibnamefont
  {Alexander}},\ }\bibfield  {title} {\enquote {\bibinfo {title} {The many-body
  reciprocal theorem and swimmer hydrodynamics},}\ }\href {\doibase
  10.1209/0295-5075/110/44001} {\bibfield  {journal} {\bibinfo  {journal}
  {{EPL}}\ }\textbf {\bibinfo {volume} {110}},\ \bibinfo {pages} {44001}
  (\bibinfo {year} {2015})}\BibitemShut {NoStop}%
\bibitem [{\citenamefont {Swan}\ and\ \citenamefont {Brady}(2007)}]{swan07}%
  \BibitemOpen
  \bibfield  {author} {\bibinfo {author} {\bibfnamefont {J.~W.}\ \bibnamefont
  {Swan}}\ and\ \bibinfo {author} {\bibfnamefont {J.~F.}\ \bibnamefont
  {Brady}},\ }\bibfield  {title} {\enquote {\bibinfo {title} {Simulation of
  hydrodynamically interacting particles near a no-slip boundary},}\ }\href
  {\doibase 10.1063/1.2803837} {\bibfield  {journal} {\bibinfo  {journal}
  {Phys. Fluids}\ }\textbf {\bibinfo {volume} {19}},\ \bibinfo {pages} {113306}
  (\bibinfo {year} {2007})}\BibitemShut {NoStop}%
\end{thebibliography}%

\end{document}